\documentclass[11pt]{article}
\usepackage{jheppub}
\usepackage[utf8]{inputenc}
\usepackage{amsmath}
\usepackage{amssymb}
\usepackage{array}
\usepackage{graphicx,xcolor,slashed}
\usepackage{enumerate}
\usepackage{ifpdf}
\usepackage{mathtools}
\usepackage{url}
\usepackage{booktabs}
\usepackage{xspace}
\usepackage{xinttools}
\usepackage{mathrsfs}
\usepackage{tensor}
\usepackage{cancel} 
\usepackage{cleveref}
\usepackage{subcaption}
\usepackage{float}
\usepackage{multirow}

\newcommand{\pol}{\ensuremath{\varepsilon}}

\newcommand{\eg}{{e.g.~}}
\newcommand{\ie}{{i.e.~}}

\DeclareMathOperator*{\cut}{Cut}
\newcommand{\cutrat}[1]{\mathcal{C}_{#1}}

\DeclareMathOperator{\sv}{sv}

\usepackage[compat=1.1.0]{tikz-feynman} \tikzstyle{every path}=[line
width=1.5pt] \tikzfeynmanset{ my dot/.style={ /tikzfeynman/dot,
    /tikz/minimum size=15pt}, }
\newcommand\de{\mathrel{\bullet\mkern-2.5mu{\rightarrow}}}
\newcommand\ue{\mathrel{\bullet\mkern-3mu{-}\mkern-3mu\bullet}}
\newcommand{\edgeCut}[5]{\coordinate (midpoint) at
  ($(#1)!0.5!(#2)$);\draw [dotted,color=#5,line width = 3pt]
  ($(midpoint) + (#3:#4)$) -- ($(midpoint) + (#3+180:#4)$);}

\usepackage{standalone}

\newcolumntype{x}[1]{>{\centering\let\newline\\\arraybackslash\hspace{0pt}}p{#1}}

\newcommand\T{\rule{0pt}{2.6ex}}       
\newcommand\B{\rule[-1.2ex]{0pt}{0pt}} 

\title{Color-kinematics dual representations of one-loop matrix elements
   in the open-superstring effective action}

\author[a,b]{Alex Edison}
\author[b]{and Micah Tegevi}

\affiliation[a]{Department of Physics and Astronomy, Northwestern University, Evanston, Illinois, 60208, USA}
\affiliation[b]{Department of Physics and Astronomy, Uppsala
  University, Box 516, 75120 Uppsala, Sweden}

\emailAdd{alexander.edison@northwestern.edu}
\preprint{UUITP-47/22}

\date{\today}

\abstract{ The $\alpha'$-expansion of string theory provides a rich
  set of higher-dimension operators, indexed by $\zeta$ values, which
  can be used to study color-kinematics duality and the double copy.
  These two powerful properties, actually first noticed in tree-level
  string amplitudes, simplify the construction of both gauge and
  gravity amplitudes.  However, their applicability and limitations
  are not fully understood.  We attempt to construct a set of
  color-kinematics dual numerators at one loop and four points for
  insertions of operator combinations corresponding to the lowest four
  $\zeta_2$-free operator insertions from the open superstring:
  $\zeta_3$, $\zeta_5$, $\zeta_3^2$, and $\zeta_7$.  We succeed in
  finding a representation for the first three in terms of box,
  triangle, and bubble numerators.  In the case of $\zeta_7$ we find
  an obstruction to a fully color-dual representation related to the
  bubble-on-external-leg type diagrams.  We discuss two paths around
  this obstruction, both of which signal an incompatability between
  color-kinematics duality and manifesting certain desired properties.
  Using the constructed color-dual numerators, we find two different
  Bern-Carrasco-Johansson double copies that produce candidate
  closed-string-insertion numerators.  Both approaches to the double
  copy match the kinematics of the cuts, with relative normalization
  set by either summing over both double copies including degeneracy
  or by including an explicit prefactor on the double-copy numerator
  definitions.}

\begin{document}
\maketitle

\section{Introduction}

The study of scattering amplitudes has long built on the interplay
between generating new theoretical data, manifesting different
properties of the data, and finding new ways of encoding the results.
One significant achievement of this progression is the discovery of
double-copy relations between gauge theory and gravity amplitudes.  At
tree level, the connections between three different manifestations of
the double copy: Kawai-Lewellen-Tye (KLT)\cite{Kawai:1985xq},
Bern-Carrasco-Johansson (BCJ)\cite{Bern:2010ue}, and Cachazo-He-Yuan
(CHY)\cite{Cachazo:2013iea}, have opened up a wide range of methods
for studying gauge and gravity amplitudes.  All three representations
can also be useful in studying loop-level amplitudes.  The KLT and CHY
formulations are effective at generating on-shell data for both gauge
and gravity theories.  The loop level incarnation of the BCJ double
copy allows calculating gravitational diagram numerators directly from
gauge theory numerators.

The tree-level BCJ double copy starts from the observation that when
writing the four-point Yang--Mills (YM) amplitude in terms of cubic
diagrams, the group Jacobi relations between the color factors of the
$s$, $t$ and $u$ channels
\begin{equation}
  c_s - c_t - c_u = 0
  \label{eq:color-jac}
\end{equation}
are a necessary relation to ensure gauge invariance of the amplitude.
If we then suppose that we can write a gravitational amplitude in
terms of the square of some numerator factor $N_g$, linearized
diffeomorphism invariance requires that
\begin{equation}
  N_s - N_t - N_u = 0 \,.
  \label{eq:kin-jac}
\end{equation}
The algebraic duality between the color Jacobi relations,
\cref{eq:color-jac}, and \cref{eq:kin-jac} is referred to as
color-kinematics (CK) duality.  The full BCJ double copy states that
if one can write an amplitude in terms of numerators that obey
\cref{eq:kin-jac} for any $s$, $t$, $u$ triplet of cubic graphs whose
color obeys \cref{eq:color-jac}, then a ``gravitational'' amplitude
can be obtained by replacing all color factors with their
corresponding numerators.  This process can even work at loop level,
with the caveat that in the color Jacobi relations one should only sum
over the color contractions within the triplet.  It has also been
shown to work for a wide range of theories.  Of interest in the
current work is that when $N_g$ is a set of numerators for an
$n$-point $\mathcal{N}=4$ super-Yang--Mills (sYM) amplitude obeying
\cref{eq:kin-jac}, then $N_g^2$ is an $n$-point numerator for
$\mathcal{N}=8$ supergravity (SG).  See Ref.~\cite{Bern:2019prr} for a
comprehensive review of the subject.

Recent decades have also seen a growing effort to explore ways that
scattering amplitudes and string theoretic methods can inform each
other.  Even just limiting the scope to the $\alpha'$-expansion of the
open superstring, we find a rich playground to: study higher-dimension
operators, generalize the notion of double copies \cite{Chi:2021mio},
find interesting representations of theories
\cite{Carrasco:2016ldy,Carrasco:2016ygv}, and even inform number
theory
\cite{Schlotterer:2012ny,Stieberger:2013wea,Stieberger:2014hba,Schlotterer:2018zce,Vanhove:2018elu,Brown:2019wna}.
Working in the other direction, amplitudes methods were recently
applied to calculating one-loop matrix elements in both the open- and
closed-string effective actions \cite{Edison:2021ebi}.  These methods
were able to verify pieces of the closed-string results of D'Hoker and
Green \cite{DHoker:2019blr}, produce new predictions for the open
string, and target terms currently unreachable by standard string
methods.

The goal of this paper is to continue building the groundwork for
exploring loop-level BCJ with higher-dimension operators.  We will
attempt to construct CK dual numerators for open-superstring matrix
elements up through $\alpha'^7$. The paper is structured as
follows. We begin by reviewing the structure of amplitudes related to
color-kinematics duality in \cref{sec:review}, including discussion on
loop-level considerations and tree-level constructions in sYM and the
open superstring. Next, in \cref{sec:cuts} we demonstrate the
simplicity of constructing one-loop unitarity cuts from
well-structured tree amplitudes, even with the higher dimension
operators of the open string low energy effective action. Finally, in
\cref{sec:cubic} we apply the method of maximal cuts
\cite{Bern:2007ct,Bern:2008pv,Bern:2010tq,Bern:2012uf,Bern:2018jmv}
in conjunction with CK duality to construct color-dual representations
for insertions of open string operators, and double copy the
numerators to find representations for closed-string operator
insertions.

\section{Review}
\label{sec:review}
We begin with a review of the structure of tree amplitudes in
(super-)Yang--Mills, with a particular eye towards the duality between
color and kinematics.  A tree-level gauge amplitude can be written
\begin{equation}
  \label{eq:treeamp}
  \mathcal{A}^{\mathrm{tree}}_n =-ig_*^{n-2}\sum_{g \in \Gamma_{3,n}}\frac{c_{g}N_{g}}{\prod_i p_{g_i}^2.},
\end{equation}
where the sum runs over the set of all cubic graphs with $n$ external
legs, $\Gamma_{3,n}$, including permutations of external legs. The
index $n$ is the number of external particles, and $g_*$ is the
coupling constant and other overall normalizations for the theory. The
kinematic denominator is the product of all propagators in the graph,
$p_{g_i}$. The numerators are decomposed into color factors $c_g$, and
kinematic factors $N_g$.  The color factors are gauge group structure
constants.  The kinematic contain Lorentz products of polarization
vectors for the external particles, $\pol_i$, as well as the momenta
for the particles, $p_i$. In this paper, we will use all-outgoing
external momenta.

We work in a normalization where the Mandelstam variables of massless
momenta carry a factor of $2$ and no explicit factors of $\alpha'$,
\ie
\begin{equation}
  s_{ij} \equiv 2 p_i \cdot p_j = (p_i + p_j)^2|_{p^2 = 0}
  \label{eq:mandelstams}
\end{equation}
with a natural extension to multi-labled $s$
\begin{equation}
  s_{ij\dots} \equiv (p_i + p_j + \dots)^2|_{p^2=0} =
  \sum_{a,b \in \{i,j,\dots\}} 2 p_a \cdot p_b \,.
\end{equation}
When off-shell momenta, like loop momenta, are included we use the
sum-of-products definition
\begin{equation}
  s_{\ell i\dots} = s_{i\dots} + \sum_{a \in \{i,\dots\}} 2 \ell \cdot p_a\,.
\end{equation}

The color factors $c_g$ come from dressing every cubic-vertex in the
graph with group theory structure constants
$f^{abc} = \mathrm{Tr}([T^a,T^b]T^c)$, where the $T^a$'s are
generators of the gauge group.  We assume that the structure constants
belong to a Lie group so that they satisfy the Jacobi relations
\begin{equation}
  f^{12\alpha}f^{\alpha 34}-f^{41\alpha}f^{\alpha 23}-f^{31\alpha}f^{\alpha 42}=0.
  \label{eq:colorjacobifs}
\end{equation}
Identifying the three products of structure constants in
\cref{eq:colorjacobifs} with the color factors for the $s$-, $t$-, and
$u$-channel cubic graphs, \cref{fig:basisTreeGraphs}, the Jacobi
relation can be simplified to
\begin{equation}
  c_s-c_t-c_u=0 \,.
  \label{eq:cjacobi}
\end{equation}
The full color-dressed amplitude $\mathcal A_n^{\mathrm{tree}}$ can be
decomposed into an $(n-2)!$ basis of color-ordered amplitudes through
judicious application of cyclicity, reflection (anti-)symmetry, the
photon decoupling identity, and the Kleiss-Kuijf relations
\cite{Mangano:1990by,Zeppenfeld:1988bz,Mangano:1988kk,Kosower:1987ic,Berends:1987cv,Cvitanovic:1980bu,Kleiss:1988ne}.
At four points, the Jacobi relation tells us that the color
coefficients of the color-ordered amplitudes can be chosen as any two
of $c_s, c_t,$ and $c_u$.

\begin{figure}
  \centering
  \raisebox{-0.5\height}{\includestandalone[scale=0.6,mode=buildnew]{figures/sDiag}}
  \qquad \qquad
  \raisebox{-0.5\height}{\includestandalone[scale=0.6,mode=buildnew]{figures/tDiag}}
  \qquad \qquad
  \raisebox{-0.5\height}{\begin{tikzpicture}
\begin{feynman}
\vertex (i1);
\vertex [above = 5 em of i1] (i2);
\vertex [below left = 2 em of i1] (e1) {\(1\)};
\vertex [above left = 2 em of i2] (e2) {\(2\)};
\vertex [right = 8 em of e2] (e3) {\(3\)};
\vertex [right = 8 em of e1] (e4) {\(4\)};
\draw (i2) -- (e4);
\draw[draw=white,double=black,line width=3pt,double distance=1.5pt] (i1) -- (e3);
\diagram{
(e1)--(i1),
(i1) --[fermion, edge label'= \(p\)] (i2),
(i2)--(e2)};
\end{feynman}
\end{tikzpicture}}
  \caption{The s-, t-, and u-channel cubic 4-point tree level graphs}
  \label{fig:basisTreeGraphs}
\end{figure}

Bern and collaborators \cite{Bern:2008qj,Bern:2010ue,Bern:2010yg}
discovered that one can arrange the kinematic numerators such that
they obey the same algebraic relations as the color factors, \eg
\begin{equation}
N\left(
\begin{gathered}

\begin{tikzpicture}
\begin{feynman}
\vertex (i1);
\vertex [right = 3 em of i1] (i2);
\vertex [below left = 2 em of i1] (e1) {\(a\)};
\vertex [above left = 2 em of i1] (e2) {\(b\)};
\vertex [above right = 2 em of i2] (e3) {\(c\)};
\vertex [below right = 2 em of i2] (e4) {\(d\)};
\diagram{
(e1)--(i1) --(e2) ,
(i1) --[fermion,color=red] (i2) -- (e4),
(i2)--(e3)};
\end{feynman}
\end{tikzpicture}
\end{gathered}\right)-
N\left(
\begin{gathered}

\begin{tikzpicture}
\begin{feynman}
\vertex (i1);
\vertex [above = 3 em of i1] (i2);
\vertex [below left = 2 em of i1] (e1) {\(a\)};
\vertex [below right = 2 em of i1] (e2) {\(d\)};
\vertex [above left = 2 em of i2] (e3) {\(b\)};
\vertex [above right = 2 em of i2] (e4) {\(c\)};
\diagram{
(e1)--(i1) --(e2) ,
(i1) --[fermion, color=red] (i2) -- (e4),
(i2)--(e3)};
\end{feynman}
\end{tikzpicture}
\end{gathered}\right)-
N\left(
\begin{gathered}

\begin{tikzpicture}
\begin{feynman}
\vertex (i1);
\vertex [above = 3.5 em of i1] (i2);
\vertex [below left = 2 em of i1] (e1) {\(a\)};
\vertex [above left = 2 em of i2] (e2) {\(b\)};
\vertex [right = 6 em of e2] (e3) {\(c\)};
\vertex [right = 6 em of e1] (e4) {\(d\)};
\draw (i2)--(e4);
\draw[draw=white,double=black,line width=3pt,double distance=1.5pt] (i1) -- (e3);
\diagram{
(i1) --[fermion, color=red] (i2),
(e1)--(i1),
(i2)--(e2)};
\end{feynman}
\end{tikzpicture}
\end{gathered}\right)=0 \,,
\label{eq:kinjacobi}
\end{equation}
for all diagrams that only differ on the four-point sub diagram.
Further, once such numerators are found, gravity amplitudes can be
constructed via a ``BCJ double copy'' \cite{Bern:2010ue} by simply
replacing the $c_g$ in \cref{eq:treeamp} with a second copy of $N_g$
\begin{equation}
  \mathcal{A}_n^{\mathrm{tree}}(1,2,\ldots,n) =\sum_{g\in \Gamma_{3,n}} \frac{c_gN_g}{D_g}
  \Rightarrow
  \mathcal{M}_n^{\mathrm{tree}}(1,2,\ldots,n)=\sum_{g\in \Gamma_{3,n}}\frac{N_g^2}{D_g} \,.
  \label{eq:bcj-dc}
\end{equation}
Since the numerator relations, \cref{eq:kinjacobi}, are algebraically
the same as the group theory Jacobi relations, they are referred to as
the kinematic Jacobi relations.  A set of numerators which obey all
possible kinematic Jacobi relations are referred to as
color-kinematics (CK) dual numerators.  Asymmetric double copies are
also possible between two distinct theories, as long as they share a
basis of cubic graphs and at least one of theories is written in terms
of CK dual numerators.  There are a wide variety of methods for
calculating manifestly CK dual numerators at tree level
\cite{Cachazo:2013hca,Cachazo:2013iea,Lam:2016tlk,Du:2017gnh,Fu:2017uzt,Du:2017kpo,Edison:2020ehu,Cheung:2021zvb,Ben-Shahar:2021zww,Carrasco:2021ptp,Carrasco:2019yyn},
one of which we will recap below in \cref{sec:amps}.  Significant work
has also gone into finding double copy representations of a wide range
of theories; we direct interested readers to Ref.~\cite{Bern:2019prr},
a recent comprehensive review of the topic.

\subsection{CK duality at one loop}
The general structure of a loop level amplitude is a straightforward
extension of the tree-level structure (dropping the coupling for
simplicity)
\begin{equation}
  \label{eq:loopamp}
  \mathcal{A}^{L\text{-loop}}_n =\sum_{g \in \Gamma^{(L)}_{3,n}}\int \left(\frac{d^D\ell}{(2 \pi)^D}\right)^L \frac{1}{S_{g}}\frac{c_{g}N_{g}}{\prod_i p_{g_i}^2.} \,,
\end{equation}
where the diagram sum is now over diagrams with exactly $L$ loops, and
the symmetry factors $S_g$ are the same ones that appear in Feynman
diagrams and take care of any over-count due to external orderings or
internal automorphisms.  Since the amplitude is again written in terms
of pairings between color and kinematic structures, color kinematics
duality can still be explored and exploited to simplify both gauge
theory and gravity calculations.  The gauge theory simplifications
come from the fact that the kinematic Jacobi relations can be used to
express all cubic loop diagram numerators in terms of a small number
of \emph{basis} diagrams.  This can be useful even in the one-loop
four-point case. The relevant graph topologies that will enter our
calculations are the box, the triangle, and the 2-2 bubble
\begin{equation}
  \Gamma^{(1)}_{3,4} = \left\{
    \begin{gathered}
      
\begin{tikzpicture}
\begin{feynman}
\path (0,0) node[vertex] (i1)
node[vertex,right = 3 em of i1] (i4)
node[vertex,above = 3 em of i1] (i2)
node[vertex,right = 3 em of i2] (i3)
node[vertex,below left = 2 em of i1,label = 235:\(1\)] (e1)
node[vertex,above left = 2 em of i2,label = 135:\(2\)] (e2)
node[vertex,above right = 2 em of i3,label = 35:\(3\)] (e3)
node[vertex,below right = 2 em of i4,label = 335:\(4\)] (e4);
\diagram{
(e1)--(i1)--(i2)--(e2),
(i2)--(i3)--(i4)--(e4),
(i3)--(e3),
(i1)--[fermion,edge label'= \(\ell\)](i4)};
\end{feynman}
\end{tikzpicture}
    \end{gathered}
    ,
    \begin{gathered}
      
\begin{tikzpicture}
\begin{feynman}
  \path (0,0) node[vertex] (i1)
  node[vertex,above = 3 em of i1] (i2)
  node[vertex,right = 3 em of i1] (i4)

  node[vertex, above = 1.4 em of i1] (t1)
  node[vertex,right = 2.5 em of t1] (i3)
  node[vertex, right = 3 em of i3] (i4)
  
  node[vertex,below left = 2 em of i1,label = 235:\(1\)] (e1)
  node[vertex,above left = 2 em of i2,label = 135:\(2\)] (e2)
  node[vertex,above right = 2 em of i4,label = 35:\(3\)] (e3)
  node[vertex,below right = 2 em of i4,label = 335:\(4\)] (e4);
\diagram{
(e1)--(i1)--(i2)--(e2),
(i2)--(i3)--(i4)--(e4),
(i4)--(e3),
(i1)--[fermion,edge label'= \(\ell\)](i3)};
\end{feynman}
\end{tikzpicture}

    \end{gathered}
    ,
    \begin{gathered}
      
\begin{tikzpicture}
\begin{feynman}
\path (0,0) node[vertex] (i1)
node[vertex, right = 1.5 em of i1] (i2)
node[vertex, right = 2 em of i2] (i3)
node[vertex,right = 1.5 em of i3] (i4)
node[vertex,below left = 2 em of i1,label = 235:\(1\)] (e1)
node[vertex,above left = 2 em of i1,label = 135:\(2\)] (e2)
node[vertex,above right = 2 em of i4,label = 35:\(3\)] (e3)
node[vertex,below right = 2 em of i4,label = 335:\(4\)] (e4);
\diagram{
(e1)--(i1)--(i2)--[half left] (i3)--(i4)--(e4),
(i4)--(e3),
(i1)--(e2),
(i3)--[anti fermion,half left,edge label= \(\ell\)](i2)};
\end{feynman}
\end{tikzpicture}
    \end{gathered}
  \right\} + \text{relabelings} \,.
  \label{eq:graph-basis}
\end{equation}
Through the kinematic Jacobi relations, we can relate the triangle and
bubble numerators to different labelings of the box numerator.  We
start by applying the Jacobi relation on the edge connecting externals
1 and 2 of the box-diagram with color-order 1234, giving the following
relation between diagrams
\begin{equation}
  \begin{gathered}
    
\begin{tikzpicture}
\begin{feynman}
\path (0,0) node[vertex] (i1)
node[vertex,right = 3 em of i1] (i4)
node[vertex,above = 3 em of i1] (i2)
node[vertex,right = 3 em of i2] (i3)
node[vertex,below left = 2 em of i1,label = 235:\(1\)] (e1)
node[vertex,above left = 2 em of i2,label = 135:\(2\)] (e2)
node[vertex,above right = 2 em of i3,label = 35:\(3\)] (e3)
node[vertex,below right = 2 em of i4,label = 335:\(4\)] (e4);
\diagram{
(e1)--(i1)--(i2)--(e2),
(i2)--(i3)--[color=red](i4)--(e4),
(i3)--(e3),
(i1)--[fermion,edge label'= \(\ell\)](i4)};
\end{feynman}
\end{tikzpicture}
  \end{gathered} - 
  \begin{gathered}
    
\begin{tikzpicture}
\begin{feynman}
\path (0,0) node[vertex] (i1)
node[vertex,right = 3 em of i1] (i4)
node[vertex,above = 3 em of i1] (i2)
node[vertex,right = 3 em of i2] (i3)
node[vertex,below left = 2 em of i1,label = 235:\(1\)] (e1)
node[vertex,above left = 2 em of i2,label = 135:\(2\)] (e2)
node[vertex,above right = 2 em of i3,label = 35:\(3\)] (e3)
node[vertex,below right = 2 em of i4,label = 335:\(4\)] (e4);
\diagram{
(e1)--(i1)--(i2)--(e2),
(i2)--(i3)--[color=red](i4)--(e4),
(i3)--(e3),
(i1)--[fermion,edge label'= \(\ell\)](i4)};
\end{feynman}
\end{tikzpicture}
  \end{gathered}- 
  \begin{gathered}
    
\begin{tikzpicture}
\begin{feynman}
  \path (0,0) node[vertex] (i1)
  node[vertex,above = 3 em of i1] (i2)
  node[vertex,right = 3 em of i1] (i4)

  node[vertex, above = 1.4 em of i1] (t1)
  node[vertex,right = 2.5 em of t1] (i3)
  node[vertex, right = 3 em of i3] (i4)
  
  node[vertex,below left = 2 em of i1,label = 235:\(1\)] (e1)
  node[vertex,above left = 2 em of i2,label = 135:\(2\)] (e2)
  node[vertex,above right = 2 em of i4,label = 35:\(3\)] (e3)
  node[vertex,below right = 2 em of i4,label = 335:\(4\)] (e4);
\diagram{
(e1)--(i1)--(i2)--(e2),
(i2)--(i3)--[color=red](i4)--(e4),
(i4)--(e3),
(i1)--[fermion,edge label'= \(\ell\)](i3)};
\end{feynman}
\end{tikzpicture}
  \end{gathered}=0.
  \label{eq:bcjloop}
\end{equation}
Expressed as the numerators of the graphs, \cref{eq:bcjloop} fixes the numerator for
the triangle graph, $N_{\bigtriangleup}(12|3|4)$, in terms of the two
box numerators
\begin{equation}
  \begin{split}
    &N_{\square}(1|2|3|4)-N_{\square}(1|2|4|3)-N_{\bigtriangleup}(1|2|3,4)
    = 0 \\ \Rightarrow
    &N_{\bigtriangleup}(1|2|3,4)=N_{\square}(1|2|3|4)-N_{\square}(1|2|4|3).
  \end{split}
  \label{eq:bcjBoxnums} 
\end{equation}
When referring to functions of loop diagrams (both numerators and
cuts), we use vertical bars to separate vertices in the loop, commas
to separate external legs which meet at a three-point vertex, and
concatenated labels for when external legs meet at a higher--point
vertex.  We can get another relation involving the 2-2 bubble diagram
by applying the Jacobi relation on the loop edge connecting externals
3 and 4 of the triangle
\begin{equation}
  \begin{gathered}
    
\begin{tikzpicture}
\begin{feynman}
  \path (0,0) node[vertex] (i1)
  node[vertex,above = 3 em of i1] (i2)
  node[vertex,right = 3 em of i1] (i4)

  node[vertex, above = 1.4 em of i1] (t1)
  node[vertex,right = 2.5 em of t1] (i3)
  node[vertex, right = 3 em of i3] (i4)
  
  node[vertex,below left = 2 em of i1,label = 235:\(1\)] (e1)
  node[vertex,above left = 2 em of i2,label = 135:\(2\)] (e2)
  node[vertex,above right = 2 em of i4,label = 35:\(3\)] (e3)
  node[vertex,below right = 2 em of i4,label = 335:\(4\)] (e4);
  \diagram{
    (i1)--[color=red](i2),
    (e1)--(i1),
    (i2)--(e2),
(i2)--(i3)--(i4)--(e4),
(i4)--(e3),
(i1)--[fermion,edge label'= \(\ell\)](i3)};
\fill (i1) circle (.75pt);
  \fill (i2) circle (.75pt);
\end{feynman}
\end{tikzpicture}
  \end{gathered}-
  \begin{gathered}
    \begin{tikzpicture}
\begin{feynman}
  \path (0,0) node[vertex] (i1)
  node[vertex,above = 3 em of i1] (i2)
  node[vertex,right = 3 em of i1] (i4)

  node[vertex, above = 1.4 em of i1] (t1)
  node[vertex,right = 2.5 em of t1] (i3)
  node[vertex, right = 3 em of i3] (i4)
  
  node[vertex,below left = 2 em of i1,label = 235:\(2\)] (e1)
  node[vertex,above left = 2 em of i2,label = 135:\(1\)] (e2)
  node[vertex,above right = 2 em of i4,label = 35:\(3\)] (e3)
  node[vertex,below right = 2 em of i4,label = 335:\(4\)] (e4);
  
  \diagram{
    (i1)--[color=red](i2),
    (e1)--(i1),
    (i2)--(e2),
    (i2)--(i3)--(i4)--(e4),
    (i4)--(e3),
    (i1)--[fermion,edge label'= \(\ell\)](i3)};
  \fill (i1) circle (.75pt);
  \fill (i2) circle (.75pt);
\end{feynman}
\end{tikzpicture}
  \end{gathered}-
  \begin{gathered}
    
\begin{tikzpicture}
\begin{feynman}
\path (0,0) node[vertex] (i1)
node[vertex, right = 1.5 em of i1] (i2)
node[vertex, right = 2 em of i2] (i3)
node[vertex,right = 1.5 em of i3] (i4)
node[vertex,below left = 2 em of i1,label = 235:\(1\)] (e1)
node[vertex,above left = 2 em of i1,label = 135:\(2\)] (e2)
node[vertex,above right = 2 em of i4,label = 35:\(3\)] (e3)
node[vertex,below right = 2 em of i4,label = 335:\(4\)] (e4);
\diagram{
  (i1)--[color=red](i2),
  (e1)--(i1),
  (i2)--[half left] (i3)--(i4)--(e4),
(i4)--(e3),
(i1)--(e2),
(i3)--[anti fermion,half left,edge label= \(\ell\)](i2)};
\end{feynman}
\end{tikzpicture}
  \end{gathered}=0.
  \label{eq:bcjloopBub}
\end{equation}
This relates the bubble numerator, $N_{\bigcirc}(1,2|3,4)$, to the
triangle and, using the relation we just found in
\cref{eq:bcjBoxnums}, the box numerators
\begin{equation}
  \begin{split}
    &N_{\bigtriangleup}(1|2|3,4)-N_{\bigtriangleup}(2|1|3,4)-N_{\bigcirc}(1,2|3,4)=0
    \\\Rightarrow
    &N_{\bigcirc}(1,2|3,4)=N_{\bigtriangleup}(1|2|3,4)-N_{\bigtriangleup}(2|1|3,4)\\&
    \qquad =
    N_{\square}(1|2|3|4)-N_{\square}(1|2|4|3)-N_{\square}(2|1|3|4)+N_{\square}(2|1|4|3).
  \end{split}
  \label{eq:bcjTriBubNum}
\end{equation}
Since \cref{eq:bcjBoxnums,eq:bcjTriBubNum} are used to define triangle
and bubble numerators in terms of the box numerator, we refer to them
as the \emph{defining} kinematic Jacobi relations.  In addition to
these defining relations, there are also \emph{boundary} kinematic
Jacobi relations that relate a difference of two non-zero diagrams to
bubble-on-external-leg (BEL) or tadpole diagrams, whose numerators we
often set to $0$.  The main boundary relation stems from applying the
Jacobi relation on one of the triangle edges only adjacent to one
external leg, which yields
\begin{equation} 
  \begin{gathered}
    
\begin{tikzpicture}
\begin{feynman}
  \path (0,0) node[vertex] (i1)
  node[vertex,above = 3 em of i1] (i2)
  node[vertex,right = 3 em of i1] (i4)

  node[vertex, above = 1.4 em of i1] (t1)
  node[vertex,right = 2.5 em of t1] (i3)
  node[vertex, right = 3 em of i3] (i4)
  
  node[vertex,below left = 2 em of i1,label = 235:\(1\)] (e1)
  node[vertex,above left = 2 em of i2,label = 135:\(2\)] (e2)
  node[vertex,above right = 2 em of i4,label = 35:\(3\)] (e3)
  node[vertex,below right = 2 em of i4,label = 335:\(4\)] (e4);
  \diagram{
    (i2)--[color=red](i3),
(e1)--(i1)--(i2)--(e2),
(i3)--(i4)--(e4),
(i4)--(e3),
(i1)--[fermion,edge label'= \(\ell\)](i3)};
\fill (i3) circle (.75pt);
  \fill (i2) circle (.75pt);
\end{feynman}
\end{tikzpicture}
  \end{gathered}
  -
  \begin{gathered}
    
\begin{tikzpicture}
\begin{feynman}
  \path (0,0) node[vertex] (i1)
  node[vertex,above = 3 em of i1] (i2)
  node[vertex,right = 3 em of i1] (i4)

  node[vertex, above = 1.4 em of i1] (t1)
  node[vertex,right = 2.5 em of t1] (i3)
  node[vertex, right = 3 em of i3] (i4)
  
  node[vertex,below left = 2 em of i1,label = 235:\(1\)] (e1)
  node[vertex,above left = 2 em of i2,label = 135:\(2\)] (e2)
  node[vertex,above right = 2 em of i4,label = 35:\(3\)] (e3)
  node[vertex,below right = 2 em of i4,label = 335:\(4\)] (e4);
  \diagram{
    (i2)--[color=red](i3),
(e1)--(i1)--[fermion,edge label= \(\ell\)](i2)--(e2),
(i3)--(i4)--(e4),
(i4)--(e3),
(i1)--(i3)};
\fill (i3) circle (.75pt);
  \fill (i2) circle (.75pt);
\end{feynman}
\end{tikzpicture}
  \end{gathered}
  -
  \begin{gathered}
    
\begin{tikzpicture}
  \begin{feynman}
    \vertex (i1);
    \vertex [below left = 4em of i1] (e1) {\(1\)};
    \vertex [above left =4em of i1] (e2) {\(2\)};
    \vertex [right=3em of i1] (i2);
    \vertex [above right=3em of i2] (e3){\(3\)};
    \vertex [below right=3em of i2] (e4){\(4\)};
    \vertex [below left=1.3em of i1] (i3);
    \vertex [below left=1.3em of i3] (i4);
    
    \diagram* {
      (i1)--[color=red](i3)--[anti fermion,half right,edge label'=\(\ell\),near start](i4)--(e1),
      (i1)--(e2),
      (i1)--(i2),
      (i2)--(e3),
      (i2)--(e4),
      (i3)--[half left](i4) };
    \fill (i1) circle (0.75pt);
\end{feynman}
\end{tikzpicture}
 
  \end{gathered} = 0 \,.
  \label{eq:bound-jac}
\end{equation}
In the ``snail'' regularization of Ref.~\cite{Bern:2012uf}, the third
diagram in the relation is either exactly zero or carries a numerator
factor of $k_1^2 \sim 0$ meaning that the kinematic Jacobi relation
reduces to a consistency condition on the triangle numerators.
Alternatively, one can use ``Minahaning'' regularization
\cite{Minahan:1987ha,Edison:2022jln}, in which the BEL numerators are not
definied to be $0$ (and can thus be defined via \cref{eq:bound-jac}),
but are regulated by relaxing momentum conservation from $n$-point to
$n+1$-point and requiring that they conspire between themselves on the
cuts and at the level of the integrand to cancel the unphysical
$k_1^{-2}$ pole.  We adopt the ``snail'' regularization in this work,
and will generally assume that the BEL numerator is identically $0$
unless explicitly stated otherwise.

While the four-point one-loop Jacobi relations are rather simple, the
same basic method can become quite complicated yet powerful for higher
loops and legs.  For example, the four-loop four-point integrand of
maximal sYM can be expressed in terms of 85 diagrams.  The numerators
for 83 of these diagrams can be expressed in terms of only two planar
diagrams, or one non-planar diagram \cite{Bern:2012uf}.  Beyond four
loops there are problems manifesting CK duality \cite{Bern:2017ucb},
and effort has been put into finding more-accessible examples for
testing the assumptions that go into finding CK numerators
\cite{Bern:2013yya,Bern:2015ooa,Mogull:2015adi,Johansson:2017bfl,Bridges:2021ebs}.

\subsection{Tree amplitudes with manifest CK duality in sYM and the
  open superstring}

\label{sec:amps}
A systematic and efficient method of constructing tree amplitudes that
manifests their double copy structures is using the Cachazo, He, and
Yuan (CHY) formalism \cite{Cachazo:2013hca}.  In this form, tree
amplitudes are given by an integral over $n$ punctures on a sphere
against a product of theory-dependent half-integrands
$\mathcal{I}_L(\sigma)$, $\mathcal{I}_{R}(\sigma)$
\begin{equation}
  \mathcal{A}_n = \int
  \frac{d^n \sigma}{\mathrm{Vol}\left(SL(2,\mathbb{C})\right)}
  \prod_a{}' \delta\left( \sum_{b \neq a} \frac{s_{ab}}{\sigma_a - \sigma_b} \right)
  \mathcal{I}_L(\sigma) \mathcal{I}_{R}(\sigma) \,.
  \label{eq:chy-int}
\end{equation}
The $\prod{}' \delta(\dots)$ fully localizes the integrals, and the
arguments of the $\delta(\dots)$ are known as the scattering
equations.  For (super-)Yang--Mills, one of the half-integrands encodes
the color information and the other half-integrand encodes the
kinematics, while for (super)gravity both half-integrands carry
kinematics.  In both cases, the integrals can be resolved in terms of
bi-adjoint scalar amplitudes $m(\alpha|\beta)$
\cite{Vaman:2010ez,Du:2011js,Bjerrum-Bohr:2012kaa,Cachazo:2013iea}
as
\begin{align}
  \mathcal{A}_{\text{sYM}}^{\text{tree}}
  &=\sum_{\sigma, \rho \in S_{n-2}}
    c\left(\begin{minipage}{0.2\textwidth}\begin{tikzpicture}
  \begin{feynman}
    \vertex (1){$1$};
    \vertex[right =of 1] (i1);
    \vertex[above =of i1] (r2){$\rho(2)$};
    \vertex[right =of i1](i2);
    \vertex[above =of i2] (r3){$\rho(3)$};
    \vertex[right =of i2](i3);
    \vertex[above =of i3] (r4){\dots};
    \vertex[right =of i3](i4);
    \vertex[above =of i4] (r5){$\rho(n-1)$};
    \vertex[right =of i4](n){$n$};
    \diagram{
      (1)--(i1)--(r2),
      (i1)--(i2)--(r3),
      (i2)--(i3)--[dashed](r4),
      (i3) --(i4)--(r5),
      (i4)--(n)};
  \end{feynman}
\end{tikzpicture}\end{minipage}\right)
    m(1, \rho, n| 1, \sigma, n)
    N\left(\begin{minipage}{0.2\textwidth}\begin{tikzpicture}
  \begin{feynman}
    \vertex (1){$1$};
    \vertex[right =of 1] (i1);
    \vertex[above =of i1] (r2){$\sigma(2)$};
    \vertex[right =of i1](i2);
    \vertex[above =of i2] (r3){$\sigma(3)$};
    \vertex[right =of i2](i3);
    \vertex[above =of i3] (r4){\dots};
    \vertex[right =of i3](i4);
    \vertex[above =of i4] (r5){$\sigma(n-1)$};
    \vertex[right =of i4](n){$n$};
    \diagram{
      (1)--(i1)--(r2),
      (i1)--(i2)--(r3),
      (i2)--(i3)--[dashed](r4),
      (i3) --(i4)--(r5),
      (i4)--(n)};
  \end{feynman}
\end{tikzpicture}\end{minipage}\right) \notag\\    
&= \sum_{\sigma, \rho \in S_{n-2}} c(1,\rho,n) m(1, \rho, n| 1, \sigma, n) N(1, \sigma, n) \label{eq:chy-ym}\\
  \mathcal{M}_{\text{GR}}^{\text{tree}}
  &=\sum_{\sigma, \rho \in S_{n-2}} \tilde{N}(1,\rho,n) m(1, \rho, n| 1, \sigma, n) N(1, \sigma, n) \label{eq:chy-gr}
\end{align}
where the sums run over permutations of $2,3, \dots , n{-}1$, the
$c(1,\rho,n)$ are DDM half-ladder color structures
\cite{DelDuca:1999rs}, and the $N$, $\tilde{N}$ are corresponding
kinematic half-ladder numerators.  These two equations manifest the
double-copy relations between gauge theory and gravity: a
gravitational amplitude can be obtained by simply replacing the color
structures of \cref{eq:chy-ym} with a second copy of the kinematic
numerator.  The (s)YM representation also manifests the decomposition
into an $(n-2)!$ basis of color-ordered amplitudes
\cite{DelDuca:1999rs,Kleiss:1988ne}, by simply restricting to a single
element in the sum over $\rho$ and stripping the color factor
\begin{equation}
  A_{\text{sYM}}^{\text{tree}}(1, \rho,n)
  = \sum_{\sigma \in S_{n-2}} m(1, \rho,n|1,\sigma,n) N(1, \sigma,n) \,.
  \label{eq:chy-coym}
\end{equation}

The $N(\beta)$ are functions of external polarizations and momenta,
and are half-ladder \emph{basis numerators} for the BCJ relations: all
non-half-ladder graph numerators can be obtained from them via
repeated application of the kinematic Jacobi relations,
\cref{eq:kinjacobi}. For example, at four points we have
\begin{align}
  A_{\text{sYM}}^{\mathrm{tree}}(1,2,3,4)&=\left(\frac{1}{s}+\frac{1}{t}\right)N(1,2,3,4)-\frac{1}{t}N(1,3,2,4) \notag\\
  &=\frac{N(1,2,3,4)}{s}+\frac{N(1,2,3,4) - N(1,3,2,4)}{t},
\end{align}
so that the $N(1,\sigma,4)$ are equivalent to the numerators in the
standard presentation of the amplitude of \cref{eq:treeamp}
\begin{align}
  N(1,2,3,4) \equiv N_s \quad N(1,3,2,4) \equiv N_u \quad N_t \equiv N(1,2,3,4) - N(1,3,2,4)\,.
\end{align}

The composition of $N(\dots)$ is dependent on the theory under
consideration.  Since the original presentation by CHY, significant
work has gone into finding efficient calculations and compact
representations of $N(\beta)$ for (super-)Yang--Mills
\cite{Mafra:2015vca,Bjerrum-Bohr:2016axv,Lam:2016tlk,Du:2017gnh,Fu:2017uzt,Du:2017kpo,Edison:2020ehu,Cheung:2021zvb}.
Most convenient for the current work is the construction developed by
one of the current authors and Teng \cite{Edison:2020ehu}, which
calculates the numerators via auxiliary combinatoric structures known
as \emph{increasing trees}.  The resulting numerators for external
gluons can be rearranged as
\begin{align}
    N(1,\sigma,n) &= \sum_{\sigma = A\cup B}\left(\pol_1\cdot
      f_{b_1}\cdot f_{b_2}\ldots f_{b_{|B|}}\cdot \pol_n\right)
    \mathfrak{N}_{\sigma}(1;A;n)
  \label{eq:NWK}
\end{align}
with linearized field strengths
$f_i^{mn} = p_i^m\pol_i^n-p_i^n\pol_i^m$, and where
$\mathfrak{N}_{\sigma}(1;A;n)$ is independent of $n$ and dependent on
the momenta of $1$ and $A$, but only the polarizations of $A$.  The
exact content of $\mathfrak{N}_{\sigma}(\dots)$ will not be relevant
for the current discussion; we will only encounter
$\mathfrak{N}_{\sigma}(a;\emptyset;b) = 1$ here.  There is also a natural
extension when particles $1$ and $n$ are fermions
\begin{equation}
  N(1_f,\sigma,n_f) = \sum_{\sigma = A\cup B}\left(\chi_1 \slashed{f}_{b_1}\ldots
    \slashed{f}_{b_{|B|}} \xi_n\right) \mathfrak{N}_{\sigma}(1;A;n)
  \label{eq:NWKf}
\end{equation}
with fermion polarizations
\begin{equation}
  \slashed{p}_i\cdot \chi_i =0, \qquad \slashed{p}_n\cdot \xi_n=\chi_n
\end{equation}
and $\gamma$-slash conventions
\begin{equation}
  \slashed{p} \equiv p^{\mu}\gamma_{\mu}, \qquad \slashed{f}_i\equiv \frac{1}{8}f_i^{\mu \nu}\left[\gamma_{\mu},\gamma_{\nu}\right]=\frac{1}{2}\slashed{p}_i\slashed{\pol}_i \label{eq:slash-conv}
\end{equation}
which is convenient for computing loops in sYM
\cite{Edison:2020uzf,Edison:2021ebi}.  With an eye towards maximal sYM
in $D=10$ and its dimensional reductions, it is most straightforward
to work in the $16 \times 16$ Weyl representation of the $10$D Dirac
matrices, in which case the fermion polarizations will be 16-component
$\text{SO}(1,9)$ spinors.  We will shorthand the products of $f$s
appearing in \cref{eq:NWK,eq:NWKf} as
\begin{align}
  &W_{2g}(1,B,n)=\varepsilon_1\cdot f_{B_1}\ldots \cdot f_{B_{|B|}}\cdot\varepsilon_n \label{eq:wG},
  \\& W_{2f}(1_f,B,n_f) = \chi_1 \slashed{f}_{B_1}\ldots \slashed{f}_{B_{|B|}} \xi_n. \label{eq:WF}
\end{align}
The particular usefulness of this representation for the task at hand
stems from the fact that the polarizations for $1$ and $n$ are
structurally distinguished from the other particles, and thus behave
well when used as the sewn legs in the construction of unitarity cuts.

The CHY integral representation for sYM, \cref{eq:chy-int}, also has
an interpretation as the $\alpha' \to 0$ limit of superstring tree
amplitudes computed via disk integrals.  We limit ourselves to a brief
presentation here, and refer the reader to Ref.~\cite{Edison:2021ebi}
and references therein for a deeper discussion of the subject. From
the disk integral representation of open-superstring amplitudes
\begin{equation}
  A_{\mathrm{super}}^{\mathrm{tree}}(1,2,\ldots,n) =
  \int\limits_{0<z_2<\ldots <z_{n-2}}\frac{dz_2dz_3\ldots
    dz_{n-2}}{\mathrm{Vol}(SL(2,\mathbb{R}))}\left\langle
    V_1(z_1)\ldots V_n(z_n)\right\rangle
\end{equation}
the worldsheet correlator $\langle V_1(z_1) \dots V_n(z_n)\rangle$ can
be rearranged with the help of total $z$ derivatives to collect all
$\alpha'$ dependence into the Koba-Nielsen factor 
\begin{align}
  A_{\mathrm{super}}^{\mathrm{tree}}(1,2,\ldots,n)
  &=
    \int\limits_{0<z_2<\ldots <z_{n-2}}\frac{dz_2dz_3\ldots
    dz_{n-2}}{\mathrm{Vol}\left(SL(2,\mathbb{R})\right)} \prod_{1 \le i <j}^{n-1} |z_i - z_j|^{2\alpha' k_i \cdot k_j}\notag \\
  &\times
    \sum_{\rho,\tau \in
    S_{n-3}}A_{\mathrm{sYM}}^{\mathrm{tree}}(1,\rho,n-1,n)
    \frac{S_0\left[\rho|\tau\right]_1}{(z_1 - z_{\tau_1}) \dots(z_n - z_{n-1})}\,,
\end{align}
where $S_0[\rho|\tau]_1$ is the field-theory KLT matrix, and is
related to the matrix inverse of $m(\alpha|\beta)$ with legs $1,n-1,n$
at fixed positions in $\alpha, \beta$
\cite{Bern:1998sv,Bjerrum-Bohr:2010pnr}.  The
$A_{\text{sYM}}^{\text{tree}}$ and $S_0[\rho|\tau]_1$ are independent
of the $z_i$.  As such, all of the $z_i$ and $\alpha'$ dependent terms can be
collected into a single object \cite{Mafra:2011nv, Broedel:2013tta}
\begin{equation}
  Z(a_1,a_2,\ldots ,a_n|b_1,b_2,\ldots ,b_n) =\!\! \int\limits_{z_{a_1}<z_{a_2}<\ldots <z_{a_n}} \!\!\! \frac{dz_1dz_2\ldots dz_n}{\mathrm{Vol}(SL_2(\mathbb{R}))}\frac{\prod_{1 \le i <j}^{n-1} |z_i - z_j|^{2\alpha' k_i \cdot k_j}}{(z_{b_1} - z_{b_2}) \dots (z_{b_n} - z_{b_1})} \,,
\end{equation}
known as $Z$ functions or $Z$-theory amplitudes.  Explicit expressions
for the $Z$-theory amplitudes can be computed using the BGap package
\cite{BGap} or extracted from the $F$ matrix representation
\cite{Schlotterer:2012ny,MZVWebsite,Broedel:2013aza}.  In the limit
$\alpha' \to 0$, $Z$-theory amplitudes become bi-adjoint amplitudes
\begin{equation}
  \lim_{\alpha' \to 0} Z(\rho|\sigma) = m(\rho|\sigma) \,.
\end{equation}
Using $Z$ functions, the open-superstring tree amplitude can be
written as
\begin{equation}
  A_{\mathrm{super}}^{\mathrm{tree}}(a_1,\ldots,a_n)=\sum_{\rho,\tau \in S_{n-3}}Z(a_1,\dots|1,\rho,n,n-1)S_0[\rho|\tau]_1A_{\mathrm{sYM}}(1,\tau,n-1,n)\,.
  \label{eq:oss-dc}
\end{equation}
The $\alpha'$ expansion of the matrix $Z \cdot S_0$ can be further
decomposed into kinematic matrices $\mathbb{P}$, $\mathbf{M}_{2n+1}$,
and $\mathbb{Q}$ \cite{Schlotterer:2012ny,MZVWebsite,Broedel:2013aza}
\begin{align}
  Z \cdot S_0
  &= \mathbb{P}\, \mathbb{Q}
    : \exp\left( \sum_{i\ge 1} \zeta_{2i+1} \mathbf{M}_{2i+1} \right):\notag \\
  &=\mathbb{P}\left(1 + \zeta_3 \mathbf{M}_3 + \zeta_5 \mathbf{M}_5+\frac{1}{2} \zeta_3^2 \mathbf{M}_3^2
    + \zeta_3 \zeta_5 \mathbf{M}_5 \mathbf{M}_3  +\zeta_{3,5}[\mathbf{M}_5,\mathbf{M}_3]+  \dots\right)
    \label{eq:os-ss}
\end{align}
which contain the kinematic information paired with $\zeta$ values.
The matrices $\mathbb{P}$ and $\mathbb{Q}$ are infinite series
containing all of the $\zeta_{2k}$ and the higher-depth
multiple-$\zeta$ values, respectively.  The entries of $\mathbf{M}_w$
are homogenous polynomials of degree $w$ in $s_{ij}$ with rational
coefficients.  The ordering colons prescribe that all products of
$\mathbf{M}_i$ resulting from the expansion of the exponential only
appear in descending order of $i$, \ie we only have
$\mathbf{M}_{i_1}\mathbf{M}_{i_2} \dots$ when $i_1>i_2>\dots$.

Through the use of \cref{eq:chy-coym} and the relation between $S_0$
and $m(\alpha|\beta)$ we can further rewrite \cref{eq:oss-dc} as
\begin{align}
  A_{\mathrm{super}}^{\mathrm{tree}}(a_1,\ldots,a_n)
  &= \!\sum_{\rho,\tau \in S_{n-3}}\!Z(a_1\dots|1,\rho,n,n{-}1) S[\rho|\tau]_1
    \!\sum_{\sigma \in S_{n-2}}\!\! m(1,\tau,n{-}1,n|1,\sigma, n) N(1,\sigma,n)\notag\\
  &= \sum_{\rho \in S_{n-2}}Z(a_1 \dots|1,\rho,n)N(1,\rho,n)\,.
    \label{eq:strTreeAmpKLT}
\end{align}
Thus we see that sYM tree amplitudes can be promoted to
open-superstring tree amplitudes by simply upgrading bi-adjoint
amplitudes to $Z$-theory amplitudes.  Further, we can interpret
\cref{eq:oss-dc} as writing the open superstring as a double copy
between the amplitudes of $Z$-theory and super-Yang--Mills.  Expanding
both sides of the double copy in $\alpha'$ leads to the schematic
identification of operators between the respective effective actions
\cite{Mafra:2016mcc}
\begin{equation}
  \begin{split}
    (\alpha')^0:F^2 &= \mathrm{sYM}\otimes \phi^3,\\
    \zeta_2(\alpha')^2:\ F^4 &=\mathrm{sYM}\otimes (\partial^2\phi^4+\phi^5),\\
    \zeta_3(\alpha')^3:\ D^2F^4+F^5 &=\mathrm{sYM}\otimes (\partial^4\phi^4+\partial^2\phi^5+\phi^6),\\
    \zeta_4(\alpha')^4:\ D^4F^4+D^2F^5+F^6 &=\mathrm{sYM}\otimes
    (\partial^6\phi^4+\partial^4\phi^5+\partial^2\phi^6+\phi^7).
  \end{split}
\end{equation}
For the first four orders in $\alpha'$, there is only one combination
of operators that appears.  Starting at $\alpha'^5$ multiple
combinations appear.  However, these operators can be grouped
according to their $\zeta$-value coefficients \cite{Kitazawa:1987xj,
  Barreiro:2005hv,Barreiro:2012aw,Koerber:2002zb, Oprisa:2005wu}.  For
instance, $\alpha'^5$ dresses two different operator combinations, one
with coefficient $\zeta_2 \zeta_3$ and another with $\zeta_5$
\cite{Schlotterer:2012ny}.  Throughout the rest of this work, we will
often refer to terms in the expansion of $Z$-theory and their
corresponding operator insertions via their $\zeta$-value coefficient,
\eg we will talk about $\zeta_3$ insertions as a shorthand for
insertions of the $D^2 F^4 + F^5$ operator combination appearing in
$Z$ theory at $\alpha'^3$.  Similarly, we will often index objects by
the $\zeta$ value of the operator they correspond to, as in
$Z_{\zeta_3}$ or $N_g^{\zeta_5}$.

Closed-string amplitudes can be expanded in a very similar manner
using the KLT relation \cite{Kawai:1985xq} which equates closed-string
amplitudes to products of open-string amplitudes via the $\alpha'$
uplift of the kinematic kernel $S_0 \to S_{\alpha'}$
\begin{equation}
  \mathcal{M}_{\text{closed}}^{\text{tree}}
  = \left(A_{\text{open}}^{\text{tree}}\right)^T \cdot S_{\alpha'} \cdot A_{\text{open}}^{\text{tree}} \,.
  \label{eq:closed-klt}
\end{equation}
Inserting \cref{eq:os-ss}, noting that the transposition operation
reverses the color ordering prescription, and with the help of
$\mathbb{P}^T S_{\alpha'} \mathbb{P} = S_0$ and
$\mathbf{M}_i^T S_0 = S_0 \mathbf{M}_i$ we get
\cite{Schlotterer:2012ny}
\begin{align}
  \mathcal{M}_{\text{closed}}^{\text{tree}}
  &= (A_{\text{sYM}}^{\text{tree}})^T S_0 \left(1 + 2\zeta_3 \mathbf{M}_3 \dots  + 2\zeta_3 \zeta_5\left( \mathbf{M}_5 \mathbf{M}_3 + \mathbf{M}_3 \mathbf{M}_5\right) + \dots\right)A_{\text{sYM}}^{\text{tree}} \,.
    \label{eq:closed-ss}
\end{align}
Making use of \cref{eq:chy-coym}, we find a multiplication of two sYM
numerator factors via the matrix kernel
$m \cdot S_0 \cdot(1 + 2\zeta_3 \mathbf{M}_3 + \dots) \cdot m$
(suppressing the various permutation labels), which can be identified
as the result of applying the single-valued map ``sv''
\cite{Schnetz:2013hqa, Brown:2013gia}\footnote{Since we only deal with
  single $\zeta$s in this work, we will omit discussion of the
  subtleties related to the ``sv'' map and multiple $\zeta$ values.}
\begin{equation}
\sv 1 = 1 \qquad \sv \zeta_{2k} = 0 \qquad \sv \zeta_{2k+1} = 2 \zeta_{2k+1}
\qquad \text{for }k \in \mathbb{N}
\label{eq:sv-def}
\end{equation}
order-by-order in $\alpha'$ on the $Z$-functions
\cite{Schlotterer:2012ny,Stieberger:2013wea,Stieberger:2014hba,Schlotterer:2018zce,Vanhove:2018elu,Brown:2019wna}
\begin{align}
  m(\gamma,|\dots) \cdot S_0 \cdot(1 + 2\zeta_3 \mathbf{M}_3 + \dots) \cdot m(\dots|\rho)
  &= \sv Z(\gamma|\rho) \,,
\end{align}
leading to a compact representation of the closed-string tree amplitude \cite{Broedel:2013tta}
\begin{equation}
  \mathcal{M}_{\text{closed}}^{\text{tree}} = \sum_{\sigma, \rho \in S_{n-2}}
  \tilde{N}(1,\rho,n) \left[\sv Z(1, \rho, n| 1, \sigma, n)\right] N(1, \sigma, n) \,.
  \label{eq:closed-chy}
\end{equation}

\section{Unitarity Cuts in bi-adjoint and Z-theory decompositions}
\label{sec:cuts}
\subsection{Open String}
As alluded to above, the packaging of tree amplitudes into a
propagator-matrix decomposition via
\cref{eq:chy-coym,eq:strTreeAmpKLT} and the further arrangement of the
numerator terms is particularly conducive to calculating one-loop
unitarity cuts, which we now demonstrate.  We restrict out attention
to four points and only external gluons in the current work.

Generically, the unitarity cut for a graph G with vertices $V(G)$ and
edge $E(G)$ is given by
\begin{equation}
  \mathcal{C}(G) = \sum_{\substack{\text{states}\\\text{over }E(G)}} \prod_{v \in V(G)} A^{\text{tree}}(v) \,.
\end{equation}
Immediately restricting to one-loop four-point, one of the most useful
cuts is the two-particle (2P) cut, which is given as
\begin{align}
  \begin{gathered}
    \begin{tikzpicture}
\begin{feynman}
\path (0,0) node[vertex] (i1)
node[vertex] at (4em,0) (i2)
node[vertex] at (2em,2em) (m1)
node[vertex] at (2em,-2em) (m2)
node[vertex,label={[xshift=1em,yshift=0.6em]\(\ell_2\)}] at (2em,1em) (cu1)
node[vertex] at (2em,3em) (cu2)
node[vertex] at (2em, -1em) (cd1)
node[vertex,label={[xshift=1em,yshift=-0.3em]\(\ell_1\)}] at (2em,-3em) (cd2)
node[below left = 3 em of i1,label = {[label distance=-1em]235:\(1\)}] (e1)
node[above left = 3 em of i1,label = {[label distance=-1em]135:\(2\)}] (e2)
node[above right = 3 em of i2,label = {[label distance=-1em]35:\(3\)}] (e3)
node[below right = 3 em of i2,label = {[label distance=-1em]335:\(4\)}] (e4);
\diagram{
(e1) -- (i1)--[quarter left](m1)--[quarter left](i2)--(e3),
(e4)--(i2)--[quarter left] (m2) --[quarter left] (i1)--(e2)};
\edgeCut{cu1}{cu2}{90}{1em}{red}
\edgeCut{cd1}{cd2}{90}{1em}{red}
\end{feynman}
\node[circle,draw,fill=gray!30,minimum size=15pt] at (0,0){};
\node[circle,draw,fill=gray!30,minimum size=15pt] at (4em,0){};
\end{tikzpicture}
  \end{gathered}&\to
  \begin{gathered}
    \begin{tikzpicture}
\begin{feynman}
\path (0,0) node[vertex] (i1)
node[vertex] at (11em,0) (i2)
node[vertex,label={[label distance=0em]0:\(-\ell_2\)}] at (2.5em,2em) (m1)
node[vertex,label={[label distance=0em]0:\(+\ell_1\)}] at (2.5em,-2em) (m2)
node[vertex,label={[label distance=0em]180:\(+\ell_2\)}] at (8.5em,2em) (m12)
node[vertex,label={[label distance=0em]180:\(-\ell_1\)}] at (8.5em,-2em) (m22)
node[below left = 3 em of i1,label = {[label distance=-1em]235:\(1\)}] (e1)
node[above left = 3 em of i1,label = {[label distance=-1em]135:\(2\)}] (e2)
node[above right = 3 em of i2,label = {[label distance=-1em]35:\(3\)}] (e3)
node[below right = 3 em of i2,label = {[label distance=-1em]335:\(4\)}] (e4);
\diagram{
(e1)--(i1)--[quarter left](m1),
(e2)--(i1)--[quarter right](m2),
(e3)--(i2)--[quarter right](m12),
(e4)--(i2)--[quarter left](m22)
};
\end{feynman}
\node[circle,draw,fill=gray!30,minimum size=15pt] at (0,0){};
\node[circle,draw,fill=gray!30,minimum size=15pt] at (11em,0){};
\end{tikzpicture}
  \end{gathered} \notag \\
\mathcal{C}(12|34)&=\sum_{\mathrm{states}}A^{\text{tree}}(\ell_1,1,2,-\ell_2)A^{\text{tree}}(\ell_2,3,4,-\ell_1)
  \label{eq:2PCutBox}
\end{align}
in the planar color ordering.  The gluonic state sums are resolved
as \cite{Roehrig:2017gbt,Kosmopoulos:2020pcd,Edison:2020uzf}
\begin{align}
  \sum_{\text{states of }\ell_i} \mathcal{C}_{\mu \nu} \pol_{\ell_i}^\mu \pol_{-\ell_i}^\nu
  + \mathcal{C}\left( \pol_{\ell_i}\cdot \pol_{-\ell_i}\right)
  \to \mathcal{C}_{\mu \nu} \eta^{\mu \nu} + (D-2) \mathcal{C}
\end{align}
where $\mathcal{C}_{\mu \nu}$ is the component of $\mathcal{C}(12|34)$
in which $\pol_{\ell_i}^\mu \pol_{-\ell_i}^\nu$ are not contracted
with each other, and $\mathcal{C}$ the one in which they are.  Note
that during the sewing of the first leg, $\mathcal{C}=0$ since
$\pol_{\ell_i}$ and $\pol_{-\ell_i}$ come from separate tree
amplitudes and thus cannot begin contracted.  However, after sewing
the first leg subsequent legs may have been contracted by previous
sewings.  The fermionic state sums are similarly
\cite{Roehrig:2017gbt,Edison:2020uzf}\footnote{This is not strictly
  true, but along with the other prescritions below properly captures
  the parity even part of the state sewing.  See \cite{Edison:2022jln} for
  a more comprehensive treatment.}
\begin{equation}
  \sum_{\text{states of }\ell_i}\chi_{\ell_i}^{\alpha}(\xi_{\ell_i})_{\beta}=\delta_{\beta}^{\alpha}.
  \label{eq:completeness}
\end{equation}
Amplitudes expressed via \cref{eq:NWK,eq:NWKf} are particularly
well-suited to applying state sums on legs $1$ and $n$, with the state
sums closing chains of $W$s into appropriate traces.  In the case of
2P cuts, we find
\begin{equation}
  \sum_{\mathrm{states}}W_{2\text{g}}(\ell_1,X,-\ell_2)W_{2\text{g}}(\ell_2,Y,-\ell_1)=\begin{cases}
    D-2 & |X| = |Y| = 0\\
    \mathrm{tr}_v(X,Y) & \text{otherwise}
  \end{cases}
\end{equation}
where
\begin{equation}
  \mathrm{tr}_v(X, Y) = f_{x_1}^\mu{}_\nu f_{x_2}^\nu{}_\rho \dots f_{y_{|Y|}}^\sigma{}_\mu \,,
\end{equation}
for sewing gluons, while for fermions
\begin{equation}
  \sum_{\mathrm{states}}W_{2f}(\ell_1,X,-\ell_2)W_{2f}(\ell_2,Y,-\ell_1)=\begin{cases}
    \mathrm{tr}_s(\mathbf{I})=2^{\frac{D}{2}-1} & |X| = |Y| = 0\\
    \mathrm{tr}_s(X,Y) & \text{otherwise}
  \end{cases}
\end{equation}
with
\begin{equation}
  \mathrm{tr}_s(X,Y) = \left( \slashed{f}_{x_1}\dots \slashed{f}_{y_1} \dots \right)^\alpha_\beta \delta^{\beta}_\alpha \,,
\end{equation}
following our Dirac matrix convetions \cref{eq:slash-conv}.
The numerator factors $\mathfrak{N}$ are unaffected other than
renaming the legs.

These trace structures were studied in detail in
Ref.~\cite{Edison:2020uzf}, and are particularly well-behaved in the
case of maximal SUSY \footnote{For ease of discussion we focus on
  $\mathcal{N}=1$ sYM in $D=10$, but as demonstrated in
  \cite{Edison:2020uzf} the various dimensional reductions with
  maximal SUSY behave identically, with a minor change in the way in
  which the cancellations occur for $|X| + |Y| = 0$.}
\begin{equation}
  \sum_{\substack{\text{states}\\s \in \{2\text{g},2f\}}}W_s(\ell_1,X,-\ell_2)W_s(\ell_2,Y,-\ell_1)=
  \begin{cases}
    0 & |X| + |Y| < 4\\
    \mathrm{tr}_v(X,Y)-\frac{1}{2}\mathrm{tr}_s(X,Y)& \text{otherwise}.
  \end{cases}
\end{equation}
The relative factor of $-\frac{1}{2}$ between the two traces is
discussed in more detail in Refs.~\cite{Edison:2020uzf,Edison:2022jln}.
Further, when $|X| + |Y|=4$ the combination of traces is the
permutation invariant contraction of four field strengths via the $t_8$-tensor:
\begin{equation}
  \mathrm{tr}(1234)-\frac{1}{2}\mathrm{tr}_s(1234)
  =\frac{1}{2} \left(\mathrm{tr}(1234)
    +\frac{1}{4}\mathrm{tr}(12)\mathrm{tr}(34)+\mathrm{cyc}(234)\right) = \frac{1}{2} t_8(1,2,3,4) \,.
  \label{eq:trt8}
\end{equation}
We will ignore the overall normalization factor of $\frac{1}{2}$ from
now on, as it will appear on all cuts calculated in this manner so can
be absorbed into the definition of the coupling constants.

Thus, we have all of the components needed to evaluate the four-point
2P cut from \cref{eq:2PCutBox} in maximal sYM
\begin{align}
  \mathcal{C}_{\text{sYM}}(12|34)
  &=\sum_{\mathrm{states}}A(\ell_1,1,2,-\ell_2)A(\ell_2,3,4,-\ell_1) \notag\\
  & =\sum_{\substack{\rho \in S_{\{1,2\}}\\\sigma \in S_{\{3,4\}}}}
  m(\ell_1,1,2,-\ell_2|\ell_1,\rho,-\ell_2)m(\ell_2,3,4,-\ell_1|\ell_2,\sigma,-\ell_1) \notag\\
  &\qquad \times \mathfrak{N}_{\rho}(\ell_1;\emptyset;-\ell_2)
    \mathfrak{N}_{\sigma}(\ell_2;\emptyset;-\ell_1)
    \sum_{\substack{\text{states}\\s \in \{2\text{g},2f\}}}W_s(\ell_1,\rho,-\ell_2)W_s(\ell_2,\sigma,-\ell_1)
\end{align}
From Ref.~\cite{Edison:2020ehu} we identify
$\mathfrak{N}(a;\emptyset;b) = 1$.  Additionally, the state sum is
resolved to $t_8(1,2,3,4)$ as discussed above, giving the well-known
result
\begin{align}
  \mathcal{C}_{\text{sYM}}(12|34)
  &= t_8(1,2,3,4)\sum_{\substack{\rho \in S_{\{1,2\}}\\\sigma \in S_{\{3,4\}}}}
  m(\ell_1,1,2,-\ell_2|\ell_1,\rho,-\ell_2)m(\ell_2,3,4,-\ell_1|\ell_2,\sigma,-\ell_1) \notag\\
  & =t_8(1,2,3,4)\left[\frac{1}{s_{\ell_11}}+\frac{1}{s_{12}}-\frac{1}{s_{12}}\right]
    \times\left[\frac{1}{s_{\ell_23}}+\frac{1}{s_{34}}-\frac{1}{s_{34}}\right]
    =\frac{t_8(1,2,3,4)}{s_{\ell_11}s_{\ell_23}}.
  \label{eq:4PT2PCut}
\end{align}
  
This construction is straightforward to extend from sYM trees to
superstring higher-dimension-operator insertions by noting that the
doubly-color-ordered amplitudes $m(\cdot |\cdot)$ are given by the
$\alpha' \to 0$ limit of $Z(\cdot | \cdot)$. Thus, to calculate a cut
for an open-superstring matrix element, all the steps in
\cref{eq:4PT2PCut} remain the same save the evaluation of the
$m(\cdot |\cdot )$ which is replaced by the relevant
$Z(\cdot | \cdot)$ expansions.  Importantly, the operator
identification via series expansion and $\zeta$ value coefficient
matching must be taken at the level of the entire cut, \eg for
$\zeta_4 \propto \zeta_2^2 \leftrightarrow D^4 F^4 + \dots$ we must
sum over $Z_1(L) Z_{\zeta_4}(R)$, $Z_{\zeta_2}(L)Z_{\zeta_2}(R)$ and
$Z_{\zeta_4}(L) Z_{1}(R)$ to get the correct cut.

For a simple example, we compute the $\zeta_2 \leftrightarrow t_8F^4$
2P cut via the above method
\begin{align}
  \mathcal{C}^{\text{open}}_{\zeta_2}(12|34)
  &= t_8(1,2,3,4) \sum_{\rho \in S_{\{1,2\}} \atop \sigma \in S_{\{3,4\}}}Z(\ell_1,1,2,-\ell_2|\ell_1,\rho,-\ell_2)Z(\ell_2,3,4,-\ell_1|\ell_2,\sigma,-\ell_1) \Big|_{\zeta_2} \notag \\
  &= t_8(1,2,3,4) \sum_{\rho \in S_{\{1,2\}} \atop \sigma \in S_{\{3,4\}}} \Big(
    m(\ell_1,1,2,-\ell_2|\ell_1,\rho,-\ell_2)
    Z_{\zeta_2}(\ell_2,3,4,-\ell_1|\ell_2,\sigma,-\ell_1)\notag \\
  &\qquad \qquad \qquad \qquad \qquad + Z_{\zeta_2}(\ell_1,1,2,-\ell_2|\ell_1,\rho,-\ell_2)
    m(\ell_2,3,4,-\ell_1|\ell_2,\sigma,-\ell_1) \Big) \notag\\
  &= - t_8(1,2,3,4) s_{12} \left( \frac{1}{s_{\ell_1 1}} + \frac{1}{s_{\ell_2 3}}\right)
\end{align}
As expected, we find that this two-particle cut supports additional residues
corresponding to triangle topology cuts, but not the two simultaneous
residues to have non-zero contribution to the box topology.

The triangle and box cuts can be extracted by taking additional
residues on the 2P cut.  The 1-3 bubble cut and one-particle cut can
be calculated in a similar manner via sewing together chains of $W_s$.
However, special care needs to be given to correct application of
momentum conservation for these cuts, first using momentum
conservation on each vertex separately to cancel potentially-singular
terms before applying the overall four-point momentum conservation
\cite{Edison:2020uzf,Edison:2021ebi,Edison:2022jln}.

\subsection{Closed string}
\label{sec:cscuts}
Gravitational cuts at one loop are nearly as simple as gauge theory
cuts.  The abstract definition of unitarity cuts via products of trees
is the same, except with gravity tree amplitudes as the constituent
objects, \eg in the 2P case
\begin{equation}
  \mathfrak{C}_{\text{grav}}(12|34) = \sum_{\text{states}} \mathcal{M}^{\text{tree}}(\ell_1,1,2,-\ell_2)
  \mathcal{M}^{\text{tree}}(\ell_2,3,4,-\ell_1) \,.
  \label{eq:grav-2p}
\end{equation}
Since gravitons are not colored, the tree amplitudes are fully
permutation invariant, and the cut is invariant under
$1 \leftrightarrow 2$ and $3 \leftrightarrow 4$.  As spin-two
particles, the state sum for gravitons is more complicated than the
one for gauge bosons \cite{Kosmopoulos:2020pcd}
\begin{align}
  &\sum_{\text{states of } \ell_i} \mathfrak{C}_{\mu \nu \rho \sigma}
    \pol_{\ell_i}^{\mu \nu}\pol_{-\ell_i}^{\rho \sigma}
    + \mathfrak{C}_{\mu \rho} \pol_{\ell_i}^{\mu \nu} \pol_{-\ell_i,\nu}^{\rho}
    + \mathfrak{C} \pol_{\ell_i}^{\mu \nu} \pol_{-\ell_i,\mu \nu} \notag \\
  &\rightarrow \frac{1}{2} \left(\mathfrak{C}^{\mu \nu}{}_{\mu \nu} +
    \mathfrak{C}^{\mu \nu}{}_{\nu \mu} -
    \frac{2}{D-2} \mathfrak{C}^{\mu}{}_{\mu}{}^{\nu}{}_{\nu}\right)
    + \frac{D(D-3)}{2(D-2)} \mathfrak{C}^\mu{}_{\mu} +
    \frac{D(D-3)}{2} \mathfrak{C} \,,
\end{align}
with $\pol^{\mu \nu}$ the graviton polarizations and the various
$\mathfrak{C}$ are pieces of the cut with the specified polarization
contractions.  Using \cref{eq:chy-gr} for the tree amplitudes and
evaluating the state sums in \cref{eq:grav-2p} for maximal
supergravity yields\footnote{Showing this result via direct state
  sewing in maximal supergravity is somewhat tedious.  However, for
  maximal supersymmetry no information is lost or corrupted by
  performing the state sums separately on each of the two sYM
  numerators $\tilde{N}$ and $N$ and then identifying
  $\tilde{\pol}_i^\mu \pol_i^\nu = \pol_i^{\mu \nu}$ as the external
  graviton polarizations \cite{Elvang:2015rqa}.}
\begin{align}
  \mathfrak{C}_{\text{max SG}}(12|34)
  &=t_8(1,2,3,4)^2 \!\!\!
    \sum_{\alpha \in S_{\{1,2\}} \atop \beta \in S_{\{3,4\}}} \!\!
    \sum_{\rho \in S_{\{1,2\}} \atop \sigma \in S_{\{3,4\}}}\!\!
    m(\ell_1,\alpha,-\ell_2|\ell_0,\rho,-\ell_2)
    m(\ell_2,\beta,-\ell_0|\ell_2,\sigma,-\ell_0) \notag \\
  &= t_8(1,2,3,4)^2 \left( \frac{1}{s_{\ell_1,1}s_{\ell_2 3}}
    + \frac{1}{s_{\ell_1,2}s_{\ell_2 3}}
    + \frac{1}{s_{\ell_1,1}s_{\ell_2 4}}
    + \frac{1}{s_{\ell_1,2}s_{\ell_2 4}} \right)
\end{align}
where we identify
$t_8(1,2,3,4)^2 = (s t A_{\text{sYM}}^{\text{tree}}(1,2,3,4))^2 = stu
M_{\text{GR}}^{\text{tree}}(1,2,3,4)$.  Similar to lifting
sYM$\to$open superstring, gravity cuts also have a natural uplift to
string-insertion cuts via replacing the gravity tree amplitudes with
closed-string ones, \cref{eq:closed-chy},
\begin{equation}
  \mathfrak{C}^{\text{closed}}(12|34) =t_8(1,2,3,4)^2 \!\!\!
  \sum_{\alpha \in S_{\{1,2\}} \atop \beta \in S_{\{3,4\}}} \!\!
  \sum_{\rho \in S_{\{1,2\}} \atop \sigma \in S_{\{3,4\}}}\!\!
  \text{sv}\Big[Z(\ell_1,\alpha,-\ell_2|\ell_1,\rho,-\ell_2)
  Z(\ell_2,\beta,-\ell_1|\ell_2,\sigma,-\ell_1)\Big] \,.
  \label{eq:direct-closed}
\end{equation}

A direct calculation of closed-string-insertion cuts is invaluable for
verifying the double-copy behavior of our CK-dual numerators.  In
particular, it is {\it a priori} unclear exactly how to apply the BCJ
double copy to get the desired closed-string insertion.  Some natural
choices are
\begin{align}
  N^{\text{closed}}_{\zeta_i \zeta_j}(g)
  &\overset{?}{\propto} N^{\text{sYM}}(g) N^{\text{open}}_{\zeta_i \zeta_j}(g) \label{eq:symxopen}\\
  &\overset{?}{\propto} N^{\text{open}}_{\zeta_i}(g) N^{\text{open}}_{\zeta_j}(g) \label{eq:openxopen}
\end{align}
or a linear combination of the two.  The first is calculationally
desirable, as at four points we have
\begin{equation}
  N^{\text{sYM}}(g) = 
  \begin{cases}
    1 & g = \text{perms of }1|2|3|4 \\
    0 & \text{otherwise}
  \end{cases}
  \Rightarrow
  N^{\text{closed}}_{\zeta_i \zeta_j}(g) \overset{?}{\propto}
  \begin{cases}
    N^{\text{open}}_{\zeta_i \zeta_j}(g) & g=\text{perms of }1|2|3|4\\
    0 & \text{otherwise,}
  \end{cases}
  \label{eq:nice-dc}
\end{equation}
\ie the closed-string insertion numerator would only be written in
terms of box diagrams.  An important thing to note is that both
\cref{eq:symxopen} and \cref{eq:openxopen} can be valid double copies,
while \cref{eq:nice-dc} implies that
\begin{equation}
  N^{\text{sYM}}(g) N^{\text{open}}_{\zeta_i \zeta_j}(g) \not \propto N^{\text{open}}_{\zeta_i}(g) N^{\text{open}}_{\zeta_j}(g) \,;
\end{equation}
they would only need to agree on the cuts such that the actual
amplitudes would only differ by a normalization.

We use ``$\propto$'' in all of these relations because the CK
construction does not care about the overall normalization of the
numerators.  A more-complete treatment may include at least a
prefactor of $\sv(\zeta_i \zeta_j)$ as part of the definition of
$N^{\text{closed}}_{\zeta_i \zeta_j}$, but then we would still be
interested if the double copy choices produce different
normalizations.  Comparing cuts constructed from
\cref{eq:symxopen,eq:openxopen} against those constructed directly via
\cref{eq:direct-closed} can identify which combinations actually
produce the desired physical observables.  We will explore this in
more detail in \cref{sec:dc-comp}, with explicit CK numerators in
hand.

\section{Construction of Color-Dual Representations}
\label{sec:cubic}
With the foundations of the theory and cut construction established,
we turn our attention to the assembly of color-kinematics-dual
representations for the odd-$\zeta$-indexed matrix elements.  From
\cref{eq:bcjBoxnums,eq:bcjTriBubNum} we know that four-point one-loop
color-kinematics-dual numerators can be written in terms of a single
basis numerator: that of the cubic box.

We will construct our color-dual representations following the method
of maximal cuts
\cite{Bern:2007ct,Bern:2008pv,Bern:2010tq,Bern:2012uf,Bern:2018jmv},
which can be briefly summarized in the current context as:
\paragraph{Method of maximal cuts:}
\begin{enumerate}
\item Assemble an ansatz for the box diagram as a homogeneous
  polynomial of momentum products. A quick analysis of how the
  $Z$-functions enter the cuts (or the contact-diagram representation
  of \cite{Edison:2021ebi}) shows that the degree of the box numerator
  must be equal to the desired $\alpha'$ order.
\item Impose power-counting constraints on the triangle and bubble
  numerators (as constructed from the box numerator via kinematic
  Jacobi relations).
\item Impose \emph{boundary Jacobi relations} on the ansatz.  These
  are the kinematic Jacobi relations which would relate triangles and
  bubbles to bubble-on-external-leg and tadpole diagrams.  They
  generally impose that an antisymmetrization around an internal edge
  is zero.
\item Impose diagram symmetries on the ansatz.  For instance, the
  triangle diagram has a $\mathbb{Z}_2 \otimes \mathbb{Z}_2$ symmetry
  corresponding to
  \begin{equation*}
    N_{\triangle}(1,2|3|4) = - N_{\triangle}(2,1|3|4) \qquad \text{and} \qquad N_{\triangle}(1,2|3|4) = - N_{\triangle}(1,2|4|3) \big|_{\ell \to \ell+k_1+k_2}
  \end{equation*}
  In the current construction, the symmetries are always a subset of the kinematic Jacobi relations.
\item Match the cuts, as calculated via \cref{sec:cuts}.  We impose
  the matching in order of ``distance from the box'': first matching
  the box itself, then the triangle cut, followed by the 2-2 (product
  of two four-point amplitudes) and 1-3 (product of three-point and
  five-point amplitude) bubbles, and finally the one-particle cut.
\end{enumerate}

Let us now turn our attention to the details of ansatz construction.
As we saw in \cref{sec:cuts}, the external state data factors out of
all cuts as an overall $t_8(1,2,3,4)$.  We require that this
factorization holds for the full integrand as well, so will leave the
$t_8$ factor implicit from now on.  We will assume that the numerators
are not secretly rational functions themselves, and thus will be
polynomials of the external Mandelstams and $k \cdot \ell$.  We
use
\begin{equation}
  \mathcal{P} = a_1s_{12} +a_2 s_{23}+a_3 s_{1\ell}+a_4 s_{2\ell}+a_5 s_{3\ell}+a_6\ell^2
  \label{eq:momBasisSum}
\end{equation}
to denote a degree one arbitrary polynomial in the variables, and
$\mathcal{P}_n$ to refer to a degree-$n$ arbitrary polynomial. For
example the degree-two ansatz $\mathcal{P}_2$ is given by
\begin{equation}
  \begin{split}
    \mathcal{P}_2 =&a_1 s_{12}^2+a_2 s_{12} s_{\text{1$\ell $}}+a_3 s_{12} s_{23}+a_4 s_{12} s_{\text{2$\ell $}}+a_5 s_{12} s_{\text{3$\ell $}}+a_6 s_{12} \ell^2+a_7 s_{\text{1$\ell $}}^2\\&+a_8 s_{\text{1$\ell $}} s_{23}+a_9 s_{\text{1$\ell $}} s_{\text{2$\ell $}}+a_{10} s_{\text{1$\ell $}} s_{\text{3$\ell $}}+a_{11} s_{\text{1$\ell $}} \ell^2+a_{12} s_{23}^2+a_{13} s_{23} s_{\text{2$\ell $}}\\&+a_{14} s_{23} s_{\text{3$\ell $}}+a_{15} s_{23}\ell^2+a_{16} s_{\text{2$\ell $}}^2+a_{17}
   s_{\text{2$\ell $}} s_{\text{3$\ell $}}+a_{18} s_{\text{2$\ell $}} \ell^2+a_{19}
   s_{\text{3$\ell $}}^2\\&+a_{20} s_{\text{3$\ell $}}\ell^2+a_{21} \ell^4.
  \end{split}
  \label{eq:momAnsTerms}
\end{equation}

From the explicit representations of the $Z$-functions
\cite{BGap,Schlotterer:2012ny,MZVWebsite}, it is clear that only sYM
supports a maximal (box) cut.  Using this observation, we can refine
the initial ansatz for the box with degree $w$ as
\begin{equation}
  N_{\square}^w = \ell_1^2\, \mathcal{P}_{w-1} + \text{cyc}(1,2,3,4)
  \label{eq:boxGenAns}
\end{equation}
where $+\text{cyc}(1,2,3,4)$ sums the various relabelings of
$\mathcal{P}_{w-1}$, as well as the four possible inverse propagators
\begin{equation}
  \ell_1^2 = (\ell+k_1)^2 \quad \ell_2^2 = (\ell+k_{12})^2
  \quad \ell_3^2 = (\ell+k_{123})^2 \quad \ell_4^2 = \ell^2 \,.
  \label{eq:boxprops}
\end{equation}
Note that this type of ansatz is in general overcomplete, as terms
like $\ell_1^2 \ell_2^4$ can be sourced from the explicit $\ell_1^2$
prefactor multiplying against the linear combination of terms from
$\mathcal{P}_{2}$ that build $\ell_2^4$, or from the explicit
$\ell_2^2$ multiplying a different linear combination from
$\mathcal{P}_{2}$ generating $\ell_1^2 \ell_2^2$.  However, this
ansatz trivializes the vanishing of the box cut as well as the cyclic
symmetry properties of the box diagram.  We will additionally impose
that the triangle and bubble ansatz have one and two powers of
$s_{12}$ respectively
\begin{equation}
N_{\bigtriangleup}(1|2|3,4)\propto s_{12} \qquad
N_{\bigcirc}(1,2|3,4) \propto s_{12}^2 \label{eq:tribubPC}
\end{equation}
so that the loop amplitude has no explicit poles in $s_{ij}$.  These
conditions are nontrivial to impose \emph{a priori} on $N_\square$,
and enter as constraints on the $a_i$.

We proceed to construct color-dual representations for the four lowest
odd-only $\zeta$ values: $\zeta_3,\zeta_5,\zeta_3^2,$ and $\zeta_7$.
For all insertions with a factor of $\zeta_2$ it should be impossible
to construct a color-dual representation due to non-trivial monodromy
relations in the open superstring
\cite{Bjerrum-Bohr:2009ulz,Stieberger:2009hq,Broedel:2012rc}.  We have
explicitly verified the construction failure for
$\zeta_2^2 \sim \zeta_4$, $\zeta_2 \zeta_3$, $\zeta_2^3$,
$\zeta_2 \zeta_5$, and $\zeta_2^2 \zeta_3$.  These constructions fail
by being unable to match $\cutrat{w}(1|2|34)$ while also satisfying
the color-kinematics relations.

\subsection[Detailed construction of
$\zeta_3 \leftrightarrow D^2 F^4 + F^5$ ]{Detailed construction
  of $\boldsymbol{\zeta_3 \leftrightarrow D^2 F^4 + F^5}$}
We begin with a detailed account of the construction of the $\zeta_3$
representation.  Examining the color-ordered cuts we find
\begin{equation}
  \label{eq:z3contacts}
  \begin{split}
    \cutrat{\zeta_3}(1|2|34)&=
     s_{12} \left(s_{12}+s_{\text{1$\ell $}}+s_{\text{2$\ell $}}+s_{\text{3$\ell $}}\right),\\
    \cutrat{\zeta_3}(12|34)&=
    \frac{s_{12}(s_{12}+s_{123\ell})}{s_{1\ell}}+\frac{s_{12}(s_{12}+s_{1\ell})}{s_{123\ell}},
    \\ \cutrat{\zeta_3}(1|234)&=
    s_{12}+s_{23}+\frac{s_{12}(s_{12}+s_{123\ell})}{s_{12\ell}}+\frac{s_{23}(s_{23}+s_{12\ell})}{s_{123\ell}}.
  \end{split}
\end{equation}
and that the one-particle cut is purely rational, thus containing no
additional information.  We can diagrammatically expand these cuts
onto cubic diagrams as
\begin{alignat}{2}
  \label{eq:z3blowupTri}
  \cutrat{}(1|2|34) &= \begin{gathered}
    
\begin{tikzpicture}
\begin{feynman}
\path (0,0) node[vertex] (i1)
node[vertex,above = 5 em of i1] (i2)
node[vertex,below left = 2 em of i1,label=235:\(1\)] (e1)
node[vertex,above left = 2 em of i2,label=135:\(2\)] (e2)
node[vertex,below right = 2.8 em of i2,my dot] (b)
node[vertex,above right= 3 em of b,label=35:\(3\)] (e3)
node[vertex,below right = 3 em of b,label=335:\(4\)] (e4);
\diagram{
(e1) --(i1) --(i2)--(e2),
(i2)--(b)-- (e3),
(b) --(e4),
(b)--[edge label = {\(\ell\)},near end,inner sep = 1pt] (i1)--(e1)};
\end{feynman}
\edgeCut{i1}{i2}{0}{0.8em}{cyan}
\edgeCut{i2}{b}{45}{0.8em}{orange}
\edgeCut{b}{i1}{135}{0.8em}{violet}
\node[circle,draw,fill=gray!30,minimum size=15pt] at (b){};
\end{tikzpicture}
  \end{gathered}&&=\begin{gathered}
    
\begin{tikzpicture}
\begin{feynman}
\path (0,0) node[vertex] (i1)
node[vertex,right = 3 em of i1] (i4)
node[vertex,above = 3 em of i1] (i2)
node[vertex,right = 3 em of i2] (i3)
node[vertex,below left = 2 em of i1,label = 235:\(1\)] (e1)
node[vertex,above left = 2 em of i2,label = 135:\(2\)] (e2)
node[vertex,above right = 2 em of i3,label = 35:\(3\)] (e3)
node[vertex,below right = 2 em of i4,label = 335:\(4\)] (e4);
\diagram{
(e1)--(i1)--(i2)--(e2),
(i2)--(i3)--(i4)--(e4),
(i3)--(e3),
(i1)--[edge label'= \(\quad \ell\)](i4)};
\edgeCut{i2}{i3}{90}{0.8em}{orange}
\edgeCut{i4}{i1}{90}{0.8em}{violet}
\edgeCut{i1}{i2}{180}{0.8em}{cyan}
\end{feynman}
\end{tikzpicture}
  \end{gathered}+
  \begin{gathered}
    
\begin{tikzpicture}
\begin{feynman}
\path (0,0) node[vertex] (i1)
node[vertex,right = 3 em of i1] (i4)
node[vertex,above = 3 em of i1] (i2)
node[vertex, above = 1.4 em of i1] (t1)
node[vertex,right = 2.5 em of t1] (i3)
node[vertex, right = 3 em of i3] (i4)
node[vertex,below left = 2 em of i1,label = 235:\(1\)] (e1)
node[vertex,above left = 2 em of i2,label = 135:\(2\)] (e2)
node[vertex,above right = 2 em of i4,label = 35:\(3\)] (e3)
node[vertex,below right = 2 em of i4,label = 335:\(4\)] (e4);
\diagram{
(e1)--(i1)--(i2)--(e2),
(i2)--(i3)--(i4)--(e4),
(i4)--(e3),
(i1)--[edge label'= \(\ell\),near start,inner sep=1pt](i3)};
\edgeCut{i1}{i2}{0}{0.7em}{cyan}
\edgeCut{i2}{i3}{45}{0.7em}{orange}
\edgeCut{i3}{i1}{135}{0.7em}{violet}
\end{feynman}
\end{tikzpicture}
  \end{gathered},\\
\label{eq:z3blowupBub}
  \cutrat{}(12|34) &= \begin{gathered}
      \begin{tikzpicture}
\begin{feynman}
\path (0,0) node[vertex] (i1)
node[vertex] at (4em,0) (i2)
node[vertex] at (2em,2em) (m1)
node[vertex] at (2em,-2em) (m2)
node[vertex] at (2em,1em) (cu1)
node[vertex] at (2em,3em) (cu2)
node[vertex] at (2em, -1em) (cd1)
node[vertex,label={[xshift=1em,yshift=-0.3em]\(\ell\)}] at (2em,-3em) (cd2)
node[below left = 3 em of i1,label = {[label distance=-1em]235:\(1\)}] (e1)
node[above left = 3 em of i1,label = {[label distance=-1em]135:\(2\)}] (e2)
node[above right = 3 em of i2,label = {[label distance=-1em]35:\(3\)}] (e3)
node[below right = 3 em of i2,label = {[label distance=-1em]335:\(4\)}] (e4);
\diagram{(e1) -- (i1)--[quarter left](m1)--[quarter left](i2)--(e3),
(e4)--(i2)--[quarter left] (m2) --[quarter left] (i1)--(e2)};
\edgeCut{cu1}{cu2}{90}{1em}{cyan}
\edgeCut{cd1}{cd2}{90}{1em}{orange}
\end{feynman}
\node[circle,draw,fill=gray!30,minimum size=15pt] at (0,0){};
\node[circle,draw,fill=gray!30,minimum size=15pt] at (4em,0){};
\end{tikzpicture}
    \end{gathered}&&= \begin{gathered}
      
\begin{tikzpicture}
\begin{feynman}
\path (0,0) node[vertex] (i1)
node[vertex,left = 3 em of i1] (i4)
node[vertex,above = 3 em of i1] (i2)
node[vertex, above = 1.4 em of i1] (t1)
node[vertex,left = 2.5 em of t1] (i3)
node[vertex, left = 3 em of i3] (i4)
node[vertex,below right = 2 em of i1,label = 335:\(4\)] (e1)
node[vertex,above right = 2 em of i2,label = 35:\(3\)] (e2)
node[vertex,above left = 2 em of i4,label = 135:\(2\)] (e3)
node[vertex,below left = 2 em of i4,label = 235:\(1\)] (e4);
\diagram{
(e1)--(i1)--(i2)--(e2),
(i2)--(i3)--(i4)--(e4),
(i4)--(e3),
(i1)--[edge label= \(\ell\),near start](i3)};
\edgeCut{i2}{i3}{135}{0.8em}{cyan}
\edgeCut{i1}{i3}{45}{0.8em}{orange}
\end{feynman}
\end{tikzpicture}
    \end{gathered}+
    \begin{gathered}
      
\begin{tikzpicture}
\begin{feynman}
\path (0,0) node[vertex] (i1)
node[vertex,right = 3 em of i1] (i4)
node[vertex,above = 3 em of i1] (i2)
node[vertex, above = 1.4 em of i1] (t1)
node[vertex,right = 2.5 em of t1] (i3)
node[vertex, right = 3 em of i3] (i4)
node[vertex,below left = 2 em of i1,label = 235:\(1\)] (e1)
node[vertex,above left = 2 em of i2,label = 135:\(2\)] (e2)
node[vertex,above right = 2 em of i4,label = 35:\(3\)] (e3)
node[vertex,below right = 2 em of i4,label = 335:\(4\)] (e4);
\diagram{
(e1)--(i1)--(i2)--(e2),
(i2)--(i3)--(i4)--(e4),
(i4)--(e3),
(i1)--[edge label'= \(\ell\),near start](i3)};
\edgeCut{i2}{i3}{45}{0.8em}{cyan}
\edgeCut{i1}{i3}{135}{0.8em}{orange}
\end{feynman}
\end{tikzpicture}
    \end{gathered}\notag \\
  &&&\qquad +
    \begin{gathered}
     
\begin{tikzpicture}
\begin{feynman}
\path (0,0) node[vertex] (i1)
node[vertex,right = 3 em of i1] (i4)
node[vertex,above = 3 em of i1] (i2)
node[vertex,right = 3 em of i2] (i3)
node[vertex,below left = 2 em of i1,label = 235:\(1\)] (e1)
node[vertex,above left = 2 em of i2,label = 135:\(2\)] (e2)
node[vertex,above right = 2 em of i3,label = 35:\(3\)] (e3)
node[vertex,below right = 2 em of i4,label = 335:\(4\)] (e4);
\diagram{
(e1)--(i1)--(i2)--(e2),
(i2)--(i3)--(i4)--(e4),
(i3)--(e3),
(i1)--[edge label'= \(\ell\),near start](i4)};
\edgeCut{i2}{i3}{90}{0.8em}{cyan}
\edgeCut{i1}{i4}{90}{0.8em}{orange}
\end{feynman}
\end{tikzpicture}
    \end{gathered}+
    \begin{gathered}
     \begin{tikzpicture}
\begin{feynman}
\path (0,0) node[vertex] (i1)
node[vertex, right = 1.5 em of i1] (i2)
node[vertex, right = 2 em of i2] (i3)
node[vertex,right = 1.5 em of i3] (i4)
node[vertex,below left = 2 em of i1,label = 235:\(1\)] (e1)
node[vertex,above left = 2 em of i1,label = 135:\(2\)] (e2)
node[vertex,above right = 2 em of i4,label = 35:\(3\)] (e3)
node[vertex,below right = 2 em of i4,label = 335:\(4\)] (e4)
node[vertex,above = 1 em of i2] (c1)
node[vertex,above = 1 em of i3] (c2)
node[vertex,below = 1 em of i2] (c3)
node[vertex,below = 1 em of i3] (c4);
\diagram{
(e1)--(i1)--(i2)--[half left] (i3)--(i4)--(e4),
(i4)--(e3),
(i1)--(e2),
(i3)--[half left,edge label= \(\qquad \ell\)](i2)};
\edgeCut{c1}{c2}{90}{0.8em}{cyan}
\edgeCut{c3}{c4}{90}{0.8em}{orange}
\end{feynman}
\end{tikzpicture}
   \end{gathered} ,\displaybreak[0]\\
  \label{eq:z3blowupBEL}
    \cutrat{}(1|234)&=  \begin{gathered}
      \begin{tikzpicture}
\begin{feynman}
\path (0,0) node[vertex] (i2)
node[vertex, left = 3.5 em of i2] (i1)
node[vertex, left = 2 em of i1,label=180:\(1\)] (e1)
node[vertex, above right = 3 em of i2, label =35:\(2\)] (e2)
node[vertex, right = 3 em of i2, label =0:\(3\)] (e3)
node[vertex, below right = 3 em of i2, label =335:\(4\)] (e4)
node[vertex,above = 1.5 em of i1] (c1)
node[vertex,above = 1.5 em of i2] (c2)
node[vertex,below = 1.5 em of i1] (c3)
node[vertex,below = 1.5 em of i2] (c4);
\diagram{
(e1)--(i1)--[half left] (i2)--(e2),
(i2)--(e3),
(i2)--(e4),
(i2)--[half left, edge label=\(\quad \ell\)] (i1)};
\edgeCut{c1}{c2}{90}{0.8em}{cyan}
\edgeCut{c3}{c4}{90}{0.8em}{orange}
\end{feynman}
\node[circle,draw,fill=gray!30,minimum size=15pt] at (0,0){};
\end{tikzpicture}
    \end{gathered}&&= \begin{gathered}
      
\begin{tikzpicture}
\begin{feynman}
\path (0,0) node[vertex] (i1)
node[vertex,right = 3 em of i1] (i4)
node[vertex,above = 3 em of i1] (i2)
node[vertex, above = 1.4 em of i1] (t1)
node[vertex,right = 2.5 em of t1] (i3)
node[vertex, right = 3 em of i3] (i4)
node[vertex,below left = 2 em of i1,label = 235:\(1\)] (e1)
node[vertex,above left = 2 em of i2,label = 135:\(2\)] (e2)
node[vertex,above right = 2 em of i4,label = 35:\(3\)] (e3)
node[vertex,below right = 2 em of i4,label = 335:\(4\)] (e4);
\diagram{
(e1)--(i1)--(i2)--(e2),
(i2)--(i3)--(i4)--(e4),
(i4)--(e3),
(i1)--[edge label'= \(\ell\),near start](i3)};
\edgeCut{i2}{i1}{0}{0.8em}{cyan}
\edgeCut{i1}{i3}{135}{0.8em}{orange}
\end{feynman}
\end{tikzpicture}
    \end{gathered}+
    \begin{gathered}
      
\begin{tikzpicture}
\begin{feynman}
\path (0,0) node[vertex] (i1)
node[vertex,right = 3 em of i1] (i4)
node[vertex,above = 3 em of i1] (i2)
node[vertex, above = 1.4 em of i1] (t1)
node[vertex,right = 2.5 em of t1] (i3)
node[vertex, right = 3 em of i3] (i4)
node[vertex,below left = 2 em of i1,label = 235:\(4\)] (e1)
node[vertex,above left = 2 em of i2,label = 135:\(1\)] (e2)
node[vertex,above right = 2 em of i4,label = 35:\(2\)] (e3)
node[vertex,below right = 2 em of i4,label = 335:\(3\)] (e4);
\diagram{
(e1)--(i1)--[edge label= \(\ell\),near start] (i2)--(e2),
(i2)--(i3)--(i4)--(e4),
(i4)--(e3),
(i1)--(i3)};
\edgeCut{i1}{i2}{0}{0.8em}{orange}
\edgeCut{i2}{i3}{45}{0.8em}{cyan}
\end{feynman}
\end{tikzpicture}
    \end{gathered}+
    \begin{gathered}
      
\begin{tikzpicture}
\begin{feynman}
\path (0,0) node[vertex] (i1)
node[vertex,right = 3 em of i1] (i4)
node[vertex,above = 3 em of i1] (i2)
node[vertex,right = 3 em of i2] (i3)
node[vertex,below left = 2 em of i1,label = 235:\(1\)] (e1)
node[vertex,above left = 2 em of i2,label = 135:\(2\)] (e2)
node[vertex,above right = 2 em of i3,label = 35:\(3\)] (e3)
node[vertex,below right = 2 em of i4,label = 335:\(4\)] (e4);
\diagram{
(e1)--(i1)--(i2)--(e2),
(i2)--(i3)--(i4)--(e4),
(i3)--(e3),
(i1)--[edge label'= \(\ell\),near start](i4)};
\edgeCut{i2}{i1}{0}{0.8em}{cyan}
\edgeCut{i1}{i4}{90}{0.8em}{orange}
\end{feynman}
\end{tikzpicture}
    \end{gathered}.
\end{alignat}
Here the dotted lines indicate cut propagators.

Noting that the explicit expression for the triangle cut in
\cref{eq:z3contacts} is a degree-2 polynomial, we see that the ansatz
for $N^{\zeta_3}_{\square}$ is covered by
\begin{equation}
  \label{eq:boxZeta3}
  N_{\square}^{\zeta_3}(1|2|3|4)=\ell_1^2\mathcal{P}_2 + \text{cyc}(1,2,3,4)
\end{equation}
exactly as in \cref{eq:momAnsTerms} with 21 free parameters,
$a_i$. Through the Jacobi relations, \cref{eq:bcjloop,eq:bcjloopBub},
we get the numerators for the cubic triangle and 2-2 bubble,
$N_{\bigtriangleup}^{\zeta_3}$ and $N_{\bigcirc}^{\zeta_3}$. Imposing
the power counting of \cref{eq:tribubPC} fixes 10 of the free
parameters, with the remaining 11 parameters only appearing in three
distinct combinations. For this case the overlap of the defining
Jacobi relations, \cref{eq:bcjBoxnums,eq:bcjTriBubNum}, and power
counting, \cref{eq:tribubPC}, turn out to be a superset of the
symmetry conditions and boundary Jacobi relations; both are trivially
satisfied by the resulting numerators. With the numerators defined we
can start imposing the cuts.

Inserting the numerator definitions into the triangle cut,
\cref{eq:z3blowupTri}, we get
\begin{align}
  \cutrat{\zeta_3}(1|2|34) &= \cut_{1;1;2}\left[
    \frac{N_{\square}^{\zeta_3}(1|2|3|4)}{\ell_3^2}+
  \frac{N_{\bigtriangleup}^{\zeta_3}(1|2|3,4)}{s_{34}}\right] \notag \\
  &= \cut_{1;1;2}\left[\frac{N_{\square}^{\zeta_3}(1|2|3|4)}{\ell_3^2}+
    \frac{N_{\square}^{\zeta_3}(1|2|3|4) {-} N_{\square}^{\zeta_3}(1|2|4|3)}{s_{34}}\right] \,,
   \label{eq:z3trieq}
\end{align}
where $\cut_{1;1;2}$ imposes $\ell_1^2 = \ell_2^2 = \ell_4^2 = 0$.
Performing the explicit matching between the left and right hand sides
of \cref{eq:z3trieq} fixes one of the remaining $a_i$.  Continuing to
the bubbles, we have
\begin{equation}
  \cutrat{\zeta_3}(12|34) = \cut_{2;2}\left[
    \frac{N_{\bigtriangleup}^{\zeta_3}(1,2|3|4)}{s_{12} \ell_1^2}
    + \frac{N_{\bigtriangleup}^{\zeta_3}(1|2|3,4)}{s_{12} \ell_3^2} + 
    \frac{N_\square^{\zeta_3}(1|2|3|4)}{\ell_1^2\ell_3^2}
    + \frac{N_{\bigcirc}^{\zeta_3}(1,2|3,4)}{s_{12}^2} \right]
\end{equation}
and
\begin{equation}
\cutrat{\zeta_3}(1|234) = \cut_{1;3}\left[
    \frac{N_{\bigtriangleup}^{\zeta_3}(1|2|3,4)}{s_{12} \ell_2^2}
    + \frac{N_{\bigtriangleup}^{\zeta_3}(1|2,3|4)}{s_{23} \ell_3^2} + 
    \frac{N_\square^{\zeta_3}(1|2|3|4)}{\ell_2^2\ell_3^2} \right]
\end{equation}
which can of course be written exclusively in terms of labelings of
the box numerator via \cref{eq:bcjBoxnums,eq:bcjTriBubNum}.  The
matching of the numerators against the cuts from \cref{eq:z3contacts}
fixes the last two $a_i$ and we have unique color-kinematics-dual
numerators for the $\zeta_3$ insertion
\begin{align}
  \label{eq:z3nums}
  N_{\square}^{\zeta_3}(1|2|3|4)=
  &-4 s_{12}^2 s_{\text{1$\ell $}}+3 s_{12}^2 s_{23}- s_{12} s_{\text{1$\ell $}} s_{23}
    +2 s_{\text{1$\ell $}}^2 s_{23}+2 s_{12} s_{23}^2
    +2 s_{\text{1$\ell $}} s_{23}^2\nonumber\\
  &-2 s_{12}^2 s_{\text{2$\ell$}}-4 s_{12} s_{\text{1$\ell $}} s_{\text{2$\ell $}}
    +2 s_{12} s_{23} s_{\text{2$\ell $}}+2 s_{23}^2 s_{\text{2$\ell $}}
    -2 s_{12} s_{\text{2$\ell $}}^2-2 s_{23} s_{\text{2$\ell $}}^2 \notag\\
  &-2 s_{12}^2s_{\text{3$\ell $}}+4 s_{12} s_{\text{1$\ell $}} s_{\text{3$\ell $}}
    -3 s_{12} s_{23} s_{\text{3$\ell $}}+4 s_{\text{1$\ell $}} s_{23} s_{\text{3$\ell $}}
    -4 s_{23} s_{\text{2$\ell $}}s_{\text{3$\ell $}}\\
  &+2 s_{12} s_{\text{3$\ell $}}^2+4 s_{12}^2 \ell^2+4 s_{12} s_{23}\ell^2
    +4 s_{23}^2 \ell^2 \,,\nonumber\\
  N_{\bigtriangleup}^{\zeta_3}(1|2|3,4)=
  & s_{12}^3-2 s_{12}^2 s_{\text{1$\ell $}}+2 s_{12}^2 s_{23}
    -3 s_{12} s_{\text{1$\ell $}} s_{23}+ s_{12}^2 s_{\text{2$\ell $}}
    +3 s_{12} s_{23} s_{\text{2$\ell $}} \notag \\
  &- s_{12}^2 s_{\text{3$\ell $}}\,,\\
  N_{\bigcirc}^{\zeta_3}(1,2|3,4)=
  &2 s_{12}^2(s_{23} - s_{13}).
\end{align}
These numerators are also provided in the attached auxiliary file.

\subsection{Higher orders}
The construction procedure for the higher $\zeta$ values remains
nearly identical to the $\zeta_3$ case, just applied to larger initial
ansatze. However, there is one major difference.  For $\zeta_3$, we
only required bubble cuts.  With the higher $\zeta$s, the $Z$-theory
power counting is such that one-particle cuts contain polynomials of
$s_{i\dots}$ in addition to rational functions, which signals a
potential contribution that is undetectable on other cuts. Thus, we
must also match the cubic numerators against this class of cuts. The
color structure of the contact tadpole is such that the two external
orderings, (1,2,3,4) and (4,3,2,1) contribute to the same color
ordering\footnote{This can be seen by considering the six-point
  half-ladder color structures \cite{DelDuca:1999rs} $c(a,1,2,3,4,b)$
  and $c(a,4,3,2,1,b)$.  Under normal circumstances these two color
  structures are distinct; attempting to convert one into the other
  eventually results in a color Jacobi identity involving a commutator
  of $a$ and $b$.  For the tadpole color structure, $a$ and $b$ are
  identified and thus commute, removing the obstruction to mapping
  them into each other.}  so to get the complete color-ordered cut we
need the expansion:
\begin{equation}
\begin{split}
\cutrat{}(1234)=&\begin{gathered}

\begin{tikzpicture}
\begin{feynman}
\path (0,0) node[vertex] (i1)
node[vertex,label = 135:\(1\)] (e1) at (-1.1,0.9)
node[vertex, label = 90:\(2\)] (e2) at (-0.4,1.2)
node[vertex,label = 25:\(3\)] (e3) at (0.4,1.2)
node[vertex, label = 65:\(4\)] (e4) at (1.1,0.9)
node[vertex, below = 3 em of i1] (lp)
node[vertex,above = 1 em of lp] (c1)
node[vertex, below = 1 em of lp, label ={[label distance =0.4em]5:\(\ell\)}] (c2);
\diagram{(i1)--[anti fermion, half left] (lp) -- [anti fermion, half left] (i1),
  (i1)--(e3), (i1)--(e4),(i1)--(e1),(i1)--(e2)};
\edgeCut{c1}{c2}{90}{0.8em}{cyan}
\end{feynman}
\node[circle,draw,fill=gray!30,minimum size=15pt] at (0,0){};
\end{tikzpicture}
\end{gathered}
+\begin{gathered}

\begin{tikzpicture}
\begin{feynman}
\path (0,0) node[vertex] (i1)
node[vertex,label = 135:\(4\)] (e1) at (-1.1,0.9)
node[vertex, label = 90:\(3\)] (e2) at (-0.4,1.2)
node[vertex,label = 25:\(2\)] (e3) at (0.4,1.2)
node[vertex, label = 65:\(1\)] (e4) at (1.1,0.9)
node[vertex, below = 3 em of i1] (lp)
node[vertex,above = 1 em of lp] (c1)
node[vertex, below = 1 em of lp, label ={[label distance =0.4em]5:\(\ell\)}] (c2);
\diagram{(i1)--[anti fermion, half left] (lp) -- [anti fermion, half left] (i1),
  (i1)--(e3), (i1)--(e4),(i1)--(e1),(i1)--(e2)};
\edgeCut{c1}{c2}{90}{0.8em}{cyan}
\end{feynman}
\node[circle,draw,fill=gray!30,minimum size=15pt] at (0,0){};
\end{tikzpicture}
\end{gathered}=
\begin{gathered}

\begin{tikzpicture}
\begin{feynman}
\path (0,0) node[vertex] (i1)
node[vertex,right = 3 em of i1] (i4)
node[vertex,above = 3 em of i1] (i2)
node[vertex, above = 1.4 em of i1] (t1)
node[vertex,right = 2.5 em of t1] (i3)
node[vertex, right = 3 em of i3] (i4)
node[vertex,below left = 2 em of i1,label = 235:\(1\)] (e1)
node[vertex,above left = 2 em of i2,label = 135:\(2\)] (e2)
node[vertex,above right = 2 em of i4,label = 35:\(3\)] (e3)
node[vertex,below right = 2 em of i4,label = 335:\(4\)] (e4);
\diagram{
(e1)--(i1)--(i2)--(e2),
(i2)--(i3)--(i4)--(e4),
(i4)--(e3),
(i1)--[edge label'= \(\ell\),near start](i3)};
\edgeCut{i1}{i3}{135}{0.8em}{cyan}
\end{feynman}
\end{tikzpicture}
\end{gathered}+
\begin{gathered}

\begin{tikzpicture}
\begin{feynman}
\path (0,0) node[vertex] (i1)
node[vertex,right = 3 em of i1] (i4)
node[vertex,above = 3 em of i1] (i2)
node[vertex, above = 1.4 em of i1] (t1)
node[vertex,right = 2.5 em of t1] (i3)
node[vertex, right = 3 em of i3] (i4)
node[vertex,below left = 2 em of i1,label = 235:\(4\)] (e1)
node[vertex,above left = 2 em of i2,label = 135:\(1\)] (e2)
node[vertex,above right = 2 em of i4,label = 35:\(2\)] (e3)
node[vertex,below right = 2 em of i4,label = 335:\(3\)] (e4);
\diagram{
(e1)--(i1)--[edge label= \(\ell\),near start](i2)--(e2),
(i2)--(i3)--(i4)--(e4),
(i4)--(e3),
(i1)--(i3)};
\edgeCut{i1}{i2}{0}{0.8em}{cyan}
\end{feynman}
\end{tikzpicture}
\end{gathered}\\
&+\begin{gathered}

\begin{tikzpicture}
\begin{feynman}
\path (0,0) node[vertex] (i1)
node[vertex,left = 3 em of i1] (i4)
node[vertex,above = 3 em of i1] (i2)
node[vertex, above = 1.4 em of i1] (t1)
node[vertex,left = 2.5 em of t1] (i3)
node[vertex, left = 3 em of i3] (i4)
node[vertex,below right = 2 em of i1,label = 335:\(4\)] (e1)
node[vertex,above right = 2 em of i2,label = 35:\(3\)] (e2)
node[vertex,above left = 2 em of i4,label = 135:\(2\)] (e3)
node[vertex,below left = 2 em of i4,label = 235:\(1\)] (e4);
\diagram{
(e1)--(i1)--(i2)--(e2),
(i2)--(i3)--(i4)--(e4),
(i4)--(e3),
(i1)--[edge label= \(\ell\),near start](i3)};
\edgeCut{i1}{i3}{45}{0.8em}{cyan}
\end{feynman}
\end{tikzpicture}
\end{gathered}
+\begin{gathered}
\begin{tikzpicture}
\begin{feynman}
\path (0,0) node[vertex] (i1)
node[vertex, right = 1.5 em of i1] (i2)
node[vertex, right = 2 em of i2] (i3)
node[vertex,right = 1.5 em of i3] (i4)
node[vertex,below left = 2 em of i1,label = 235:\(1\)] (e1)
node[vertex,above left = 2 em of i1,label = 135:\(2\)] (e2)
node[vertex,above right = 2 em of i4,label = 35:\(3\)] (e3)
node[vertex,below right = 2 em of i4,label = 335:\(4\)] (e4)
node[vertex,above = 1 em of i2] (c1)
node[vertex,above = 1 em of i3] (c2)
node[vertex,below = 1 em of i2] (c3)
node[vertex,below = 1 em of i3] (c4);
\diagram{
(e1)--(i1)--(i2)--[half left] (i3)--(i4)--(e4),
(i4)--(e3),
(i1)--(e2),
(i3)--[half left,edge label= \(\qquad \ell\)](i2)};
\edgeCut{c3}{c4}{90}{0.8em}{cyan}
\end{feynman}
\end{tikzpicture}
\end{gathered}+
\begin{gathered}

\begin{tikzpicture}
\begin{feynman}
\path (0,0) node[vertex] (i1)
node[vertex,right = 3 em of i1] (i4)
node[vertex,above = 3 em of i1] (i2)
node[vertex,right = 3 em of i2] (i3)
node[vertex,below left = 2 em of i1,label = 235:\(1\)] (e1)
node[vertex,above left = 2 em of i2,label = 135:\(2\)] (e2)
node[vertex,above right = 2 em of i3,label = 35:\(3\)] (e3)
node[vertex,below right = 2 em of i4,label = 335:\(4\)] (e4);
\diagram{
(e1)--(i1)--(i2)--(e2),
(i2)--(i3)--(i4)--(e4),
(i3)--(e3),
(i1)--[edge label'= \(\ell\),near start](i4)};
\edgeCut{i4}{i1}{90}{0.8em}{cyan}
\end{feynman}
\end{tikzpicture}
\end{gathered}+(\ell \leftrightarrow -\ell) \,.
\end{split}
\end{equation}
At $(\alpha')^{\ge 5}$ the boundary Jacobi relations are no longer a
subset of the defining Jacobi relations and thus we need to impose
their constraints on the numerator.  We find that the symmetries are
automatically satisfied after application of all kinematic Jacobi
relations including the boundary relations.

Beyond the need to actually impose boundary Jacobi relations, the
$\zeta_5$ and $\zeta_3^2$ constructions follow exactly the same
process as the $\zeta_3$ construction.  The triangle and bubble cuts
impose a small number of conditions on the ansatz, after which the
one-particle cuts are satisfied without imposing additional
constraints.  Unlike the $\zeta_3$ numerator, the $\zeta_5$ and
$\zeta_3^2$ numerators are not unique.  After all conditions are
imposed, the $\zeta_5$ numerator has two free parameters remaining,
while the $\zeta_3^2$ numerator has only one.

At $\zeta_7$ our construction fails.  The first point of trouble
occurs in the matching of $\cutrat{\zeta_7}(12|34)$ and
$\cutrat{\zeta_7}(1|234)$.  Each cut can be matched separately, but
the two cannot be matched simultaneously.  This difficulty can be
surmounted by introducing a non-trivial numerator for the
bubble-on-external-leg (BEL) cubic diagram (the third diagram in
\cref{eq:bound-jac}). In the ``snail'' regularization
\cite{Bern:2012uf} the singular $\frac{1}{k_1^2}$ propagator that
would be expected in such a diagram is allowed to be formally non-zero
at the level of the integrand, and is then cancelled by introducing an
overall factor of $k_1^2$ in the diagram's numerator.
This procedure effectively introduces a contact correction with the
problematic propagator collapsed, carrying only the color structure of
the cubic diagram.  After double copying to gravitational theories,
these new diagrams will drop out from the representation as the
double-copied numerator will contain a $k_1^4$ while there will only
be one factor of $\frac{1}{k_1^2}$ as a propagator, leaving a $k_1^2$
which can be safely set to zero using on-shell conditions.

While this process solves the BEL obstruction, problems continue to
arise when attempting to match the one-particle cut.  Unlike in the
three previous cases, the local terms of the one-particle cut are not
automatically matched by the constrained numerators.  By itself this
is not a problem.  Unfortunately, we further find that the matching
cannot be completed with the remaining freedom in the ansatz.  As with
the BEL numerators, one might hope to introduce only
appropriately-regularized cubic tadpole contributions that will drop
out of the resulting double-copied theory.  Our attempts to absorb the
necessary local contributions into symmetry-obeying cubic tadpoles
have failed.  However, there are significantly more subtleties with
introducing cubic tadpole numerators to absorb this mismatch, so we
cannot claim that finding appropriate cubic tadpoles is impossible.
Even though the single-copy numerators do not match all cuts, since
they match the triangle and bubble cuts and obey color-kinematics
duality they will compute the triangle and bubble cuts correctly in
the double-copy theory.

\begin{table}[h!]
\small
\centering
\def\arraystretch{1.5}
\begin{tabular}{|p{5em}|x{5em}|x{8em}|x{7em}|x{5em}|}
\hline
  \textbf{$\boldsymbol{\zeta}$ \textbf{weight}}
  &\textbf{Initial ansatz}
  &\textbf{After powercounting}
  &\textbf{After boundary Jacobies}
  &\textbf{Free params after cuts}\\
\hline
  $\zeta_5$ & 126 & 43 & 25 & 2\\
  \hline
  $\zeta_3^2$ & 252 & 135& 45&1 \\
  \hline
  \multirow{2}{*}{$\zeta_7$} & 462 (+30) &223 (+30) &118 (+30)&-\\
  & 462 &352 $\left(\substack{\text{relaxed}\\\text{bubble}}\right)$ &280 & 4\\
  \hline
\end{tabular}
\caption{Number of free parameters for $\zeta_5$, $\zeta_3^2$ and
  $\zeta_7$ at each step in the construction.  The exact number of
  parameters at each step depends on the representation of the initial
  ansatz.  The final number of free parameters is independent on such
  choices.  For $\zeta_7$, we report on two methods.  First we use an
  additional 30 parameter ansatz for the bubble-on-external-leg
  numerator which is unaffected by powercounting and Jacobi relations.
  Since this construction of $\zeta_7$ is incomplete, we do not
  present the final free parameter count.  Second, we relax the bubble
  powercounting constraint which allows a complete solution.}
\label{tab:z5eqs}
\end{table}

Note that the above problems only arise when trying to force the
representation to contain strictly cubic diagrams.  The $\zeta_7$
representation from Ref.~\cite{Edison:2021ebi} does not suffer from
any of these problems, but is a purely contact representation with no
cubic diagrams.  It is possible that a full ``Minahaning''
\cite{Minahan:1987ha,Edison:2022jln} treatment of the BEL numerator
would alleviate the problems, but the required computational overhead
for that approach is prohibitive for the current work.

On the other hand, rather than modifying or lifting the boundary
Jacobi conditions, we could interrogate the necessity of the
power-counting constraints, \cref{eq:tribubPC}.  These constraints are
imposed to manifest the property that the amplitude has no poles in
external Mandelstams, but strictly speaking this only needs to hold
after integration.  Since the construction obstacle occurs at the
level of the bubble cuts, we loosen the powercounting restrictions on
the bubble to
\begin{equation}
  N_{\bigcirc}(1,2|3,4) \propto s_{12} \,.
\end{equation}
Repeating the construction with the weaker powercounting succeeds,
with four free parameters left over after all conditions are imposed.

The number of free parameters at each step in the construction for
each of $\zeta_5,\ \zeta_3^2,\ \zeta_7$ are summarized in
\cref{tab:z5eqs}. Since the degree of the box numerator is the same as
the $\alpha'$ order, the resulting numerators are sufficiently large
that they are not worth presenting in text.  They may be found in the
auxiliary file for the arXiv posting, with the remaining free
parameters left unfixed.

Color-kinematics representations of numerators are often also manifest
loop power-counting representations, isolating the leading ultraviolet
divergences to particular diagram classes.  For instance, there are
two different representations for the four-loop four-point intergrand
in maximal super-Yang--Mills: one without CK \cite{Bern:2010tq}, and
one with manifest CK \cite{Bern:2012uf}.  The one with manifest CK
numerators also manifests the UV structure of the theory by pushing
the bad power-counting terms from the box-like topolgies into the
bubble-like ones.  The UV critical dimension and leading divergence
can be extracted from only these bubble-like numerators.

We can ask the same questions of our new representations (only using
the $\zeta_7$ with the relaxed bubble power counting).  Namely, how
does the CK representation assign loop momentum power counting, and
does the CK dual representation generate the leading UV divergence
from only the bubble?  In \cref{tab:uv-pc} we present the relevant
data.  We find that the $\zeta_3$, $\zeta_5$, and $\zeta_7$
representations \emph{do} have manifest UV properties.  The scalar
bubble integral first develops a UV divergence in $D_C = 4$, and the
bubble numerator for $\zeta_5$ ($\zeta_7$) carries two (four) additional
powers of $\ell$ to produce its critical dimension $D_C=2\, (0)$.
Interestingly, while the $\zeta_3^2$ representation does not have
manifest UV behavior, the two powers of $\ell$ enter the bubble
numerator in such a way that the leading UV divergence cancels at the
level of the individual diagram.  Note that while this is not the best
imaginable power counting, it is significantly closer than the contact
representation from Ref.~\cite{Edison:2021ebi}.  That representation
also carries $\ell^2$ on bubble diagrams, but in addition it contains
explicit tadpole numerators, thus requiring a careful interplay
between multiple diagrams to see the UV cancellations.

\begin{table}[h]
\small
\centering
\def\arraystretch{1.5}
\begin{tabular}{|p{5em}|x{5em}|x{8em}|x{7em}|x{7em}|}
\hline
  \textbf{$\boldsymbol{\zeta}$ \textbf{weight}}
  &\textbf{Box }$\ell$
  &\textbf{Triangle }$\ell$
  &\textbf{2-2 Bubble }$\ell$
  &\textbf{UV Crit. Dim.}\\
\hline
  $\zeta_3$ & $\ell^2$ & $\ell$ & 1&4\\
  \hline
  $\zeta_5$ & $\ell^4$ & $\ell^3$& $\ell^2$&2\\
  \hline
  $\zeta_3^2$ &$\ell^4$ & $\ell^3$& $\ell^2$&4\\ \hline
  $\zeta_7\, \left(\substack{\text{relaxed}\\\text{bubble}}\right)$&
        $\ell^6$ & $\ell^5$ & $\ell^4$ & 0 \\                  
  \hline
\end{tabular}
\caption{The loop momentum power counting for each of the topologies,
  and the lowest dimension in which the representation develops a UV
  divergence.  The critical dimensions are in agreement with those
  reported in Ref.~\cite{Edison:2021ebi}.}
\label{tab:uv-pc}
\end{table}

\subsection{Double-copy cut comparison}
\label{sec:dc-comp}
With color-dual numerators in hand, we can easily construct a large
number of closed-string-insertion matrix elements.  In particular, the
double copy gives easy access to all closed-string matrix elements in
$\{\text{sYM},\zeta_3,\zeta_5,\zeta_3^2,\zeta_7\} \otimes
\{\text{sYM},\zeta_3,\zeta_5,\zeta_3^2,\zeta_7\}$.  Using the direct
cut construction method from \cref{sec:cscuts}, we can verify the
double-copy integrands and explore the different methods of arriving
at representations for products of $\zeta$ values.  In
\cref{tab:dc-compare}, we present the various ratios between cuts
constructed by taking double copies of our representations and cuts
calculated directly from \cref{eq:direct-closed}, without adopting any
shifts in normalization.
\begin{table}[h]
  \begin{center}
  \begin{tabular}{|c|c|c|}
    \hline
    Double-copy form
    & $\frac{\text{double-copy cuts}}{\text{direct cuts}}$
    & Analogous tree-level structure \T\B\\\hline
    $\text{sYM} \otimes \zeta_3$ & $\frac{1}{2}$ & $\mathbf{M}_{3,L}+ \mathbf{M}_{3,R}$ \T\B\\\hline
    $\text{sYM} \otimes \zeta_5$ & $\frac{1}{2}$ & $\mathbf{M}_{5,L}+ \mathbf{M}_{5,R}$ \T\B\\\hline
    $\text{sYM} \otimes \zeta_7$ & $\frac{1}{2}$ & $\mathbf{M}_{7,L} + \mathbf{M}_{7,R}$ \T\B\\\hline
    $\zeta_3 \otimes \zeta_3$ & $\frac{1}{2}$ &$\frac{1}{2}( 2\mathbf{M}_{3,L} \mathbf{M}_{3,R})$\T\\
    $\text{sYM} \otimes \zeta_3^2$ & $\frac{1}{4}$ & $\frac{1}{2}( \mathbf{M}_{3,L}^2+ \mathbf{M}_{3,R}^2)$ \T\B\\\hline
    $\zeta_3 \otimes \zeta_5$ & $\frac{1}{4}$ & $\mathbf{M}_{3,L}\mathbf{M}_{5,R} + \mathbf{M}_{5,L}\mathbf{M}_{3,R}$\T\\
    $\text{sYM} \otimes \zeta_3 \zeta_5$ & conjecture: $\frac{1}{4}$ & $\mathbf{M}_{5,L}\mathbf{M}_{3,L} + \mathbf{M}_{3,R}\mathbf{M}_{5,R}$\T\B\\\hline
    $\zeta_5 \otimes \zeta_5$ & $\frac{1}{2}$ & $\frac{1}{2} ( 2\mathbf{M}_{5,L} \mathbf{M}_{5,R})$\T\\
    $\text{sYM} \otimes \zeta_5^2$ & conjecture: $\frac{1}{4}$ & $\frac{1}{2} (\mathbf{M}_{5,L}^2 + \mathbf{M}_{5,R}^2)$\T\B\\\hline
    $\zeta_3 \otimes \zeta_7$ & $\frac{1}{4}^*$ & $\mathbf{M}_{3,L}\mathbf{M}_{7,R} + \mathbf{M}_{7,L}\mathbf{M}_{3,R}$\T\\
    $\text{sYM} \otimes \zeta_3\zeta_7$ & conjecture: $\frac{1}{4}$ & $\frac{1}{2} (\mathbf{M}_{7,L}\mathbf{M}_{3,L} + \mathbf{M}_{3,R}\mathbf{M}_{7,R})$\T\B\\\hline
    $\zeta_5 \otimes \zeta_7$ & $\frac{1}{4}^*$ & $\mathbf{M}_{5,L}\mathbf{M}_{7,R} + \mathbf{M}_{7,L}\mathbf{M}_{5,R}$\T\\
    $\text{sYM} \otimes \zeta_5 \zeta_7$ & conjecture: $\frac{1}{4}$ & $\frac{1}{2} (\mathbf{M}_{7,L}\mathbf{M}_{5,L} + \mathbf{M}_{5,R}\mathbf{M}_{7,R})$\T\B\\\hline
    $\zeta_7 \otimes \zeta_7$ & $\frac{1}{2}^*$ & $\frac{1}{2} ( 2\mathbf{M}_{7,L} \mathbf{M}_{7,R})$\T\\
    $\text{sYM} \otimes \zeta_7^2$ & conjecture: $\frac{1}{4}$ & $\frac{1}{2} (\mathbf{M}_{7,L}^2 + \mathbf{M}_{7,R}^2)$ \B\\\hline
  \end{tabular}
  \caption{A comparison between the various BCJ double-copy
    constructions buildable from sYM, $\zeta_3$, $\zeta_5$,
    $\zeta_3^2$, and $\zeta_7$ representations.  The first column
    lists the possible constructions, and $\otimes$ should be
    understood as commutative.  The second column highlights the
    differences between the symmetric and asymemtric constructions.
    While we do not have access to both forms of double-copy
    numerators for $\alpha{}'^{\ge8}$, we conjecture that if such
    numerators exist then the normalization pattern we observe in the
    asymmetric double copy continues.  The third column recalls
    analogous tree-level kinematic structures appearing in the KLT
    representation of closed-string amplitudes, \cref{eq:closed-ss}.
    The $\zeta_7$ entries used the relaxed-bubble-powercounting
    representation.  The starred entries in the second column were
    only checked on the triangle cut, $\mathfrak{C}(1|2|34)$.}
  \label{tab:dc-compare}
\end{center}
\end{table}
We find that both double copy prescriptions,
\cref{eq:symxopen,eq:openxopen}, reproduce the correct kinematic
structure of the cuts but lead to the wrong overall normalization.

To account for the normalization mismatch, we need to dig a bit deeper
into what the various double-copy constructions are doing.  First,
note that the BCJ double copy, \cref{eq:bcj-dc}, is generally
commutative: for a product between two theories it does not matter
which is assigned to $N$ and which to $\tilde{N}$.  As such the
notation $X \otimes Y$ in the table and double-copy numerator
definitions in \cref{eq:symxopen,eq:openxopen} do not distinguish
between $N_X \tilde{N}_Y$ or $N_Y \tilde{N}_X$, and so have no reason
to sum over the contributions.  On the other hand, the KLT double
copy used to define the direct cut construction,
\cref{eq:closed-klt,eq:closed-ss,eq:closed-chy,eq:direct-closed}, sum
over contributions to each $\zeta$ value from both the ``left'' theory
and the ``right'' theory.  We can make this clear by artificially
breaking the symmetry between the ``left'' and ``right'' copies of the
open string by putting ``L'' and ``R'' tag labels on
\cref{eq:closed-klt},
\begin{equation}
\mathcal{M}^{\text{closed}}
= (A_{\text{open},L})^T S_{\alpha'} A_{\text{open},R}
\end{equation}
and following the same expansion and commutation process, yielding 
\begin{equation}
  \mathcal{M}^{\text{closed}}= A_{\text{sYM}} S_0 \left( 1 + \zeta_3(\mathbf{M}_{3,L} + \mathbf{M}_{3,R}) + \dots\right) A_{\text{sYM}} \,.
  \label{eq:leftrightdc}
\end{equation}
Clearly the closed-string tree amplitude sees contributions from both
the ``left'' and ``right'' theories, even though they are identical.
Thus, in matching the cuts using BCJ double-copy numerators, we should
really be summing over all relevant combinations of $N$ and
$\tilde{N}$, even those that look degenerate.  For example, consider
the $\zeta_3^2$ contributions, which on all cuts behave like
\begin{equation}
  \mathfrak{C}_{\zeta_3^2 \otimes \text{sYM}}+
  \mathfrak{C}_{\text{sYM}\otimes \zeta_3^2} + 
  \mathfrak{C}_{\zeta_3 \otimes \zeta_3} =
  \mathfrak{C}_{\zeta_3^2}\left( \frac{1}{4} + \frac{1}{4} + \frac{1}{2}\right) =
  \mathfrak{C}_{\zeta_3^2} \,.
\end{equation}
We list the relevant terms in the expansion of \cref{eq:leftrightdc}
in the right-most column of \cref{tab:dc-compare}.  Through this lens,
the relative combinatorics is mundane while the amazing thing is that
both of the double copy forms correctly reproduce the kinematics of
the cuts individually.  Cut-matching could have instead required
cancellations between the different double-copy numerators.  We
conjecture that this normalization pattern continues for higher
$\alpha{}'$ operators.  Due to the difficulties with the $\zeta_7$
numerators mentioned above, we limit ourselves to the more-symmetric
double copy numerators which do continue to follow the observed
pattern.  It is also interesting that adopting an additional
normalization on $N^{\text{closed}}_{\zeta_i \zeta_j}$ of
$\sv(\zeta_i \zeta_j)$ leads to an exact match for all of the
asymmetric double-copy numerators, and a degeneracy overcount for the
symmetric numerators.

\section{Conclusions}
In this work, we constructed new color-dual representations at one
loop and four points that contain operator insertions from the
open-string low-energy effective action.  The three lowest mass
dimension operators for which this was possible, $D^2 F^4 + F^5$ (with
coefficient $\zeta_3 \alpha'^3$), $D^6F^4 + \dots$(with coefficient
$\alpha'^5 \zeta_5$), and $D^8 F^4 + \dots$(with coefficient
$\alpha'^6 \zeta_3^2$) , are represented in terms of box, triangle,
and 2-2 bubble numerators.  Via the kinematic Jacobi relations, the
latter two numerators are expressed purely in terms of linear
combinations of box numerators with different labelings.  For the next
mass dimension operator, the $D^{10} F^4 + \dots$ combination linked
with $\alpha'^7 \zeta_7$, we introduced a bubble-on-external-leg
contribution via ``snail'' regularization to match both bubble cuts,
and could not fully match the one-particle cuts.  Interestingly, when
double copied with sYM this operator becomes a $D^8 R^4$ type
operator, the same type which was found to cause a deviation between
the $\mathcal{N}=4$ sYM and $\mathcal{N}=8$ SUGRA UV behavior at five
loops \cite{Bern:2018jmv}.  The known five-loop sYM representations
also have unresolved tension between color-kinematics duality and
unitarity constraints \cite{Bern:2017ucb,Bern:2018jmv}.  We saw that
it was possible to overcome the $\zeta_7$ construction obtacle using a
relaxed powercounting condition on the bubble.  It is also possible
that the failure of the $\zeta_7$ construction may be overcome by
adopting the less-restrictive ``Minahan''
\cite{Minahan:1987ha,Edison:2022jln} regularization for BEL diagrams.
Both approaches provide guidance for a re-examination of the five loop
construction with considerations of relaxing either the ``snail''
scheme or diagrammatic powercounting restrictions related to forced powers of external mandelstams.

The $\zeta_3$, $\zeta_5$, and $\zeta_7$ representations have manfiest
UV behavior, with the critical dimension and actual divergence coming
from only the bubble numerators.  On the other hand, the $\zeta_3^2$
representation does not have manifest power counting.  The UV critical
dimension of $D_C=4$ suggests that the numerators should at worst be a
tensor triangle or scalar bubble. Instead, the CK dual numerator picks
an $\ell^2$ bubble.  However, the cancellation of the potential
$D_C=2$ UV divergence happens at the level of an individual bubble
diagram, and does not require summing over separate diagrams or
channels, in contrast to the contact representation of
Ref.~\cite{Edison:2021ebi}.

We hope that the representations we constructed will be useful for
studying loop-level color-kinematics duality and BCJ relations in a
simpler context than five-loop sYM or two- and three-loop
$\mathcal{N}=0,1,2$ YM.  In particular, it would be interesting to
investigate if the tree-level kinematic composition construction of
Carrasco, Rodina, Yin, and Zekioglu
\cite{Carrasco:2021ptp,Carrasco:2019yyn} can be lifted to loop level,
for which our representations could serve as seed data.

As an additional bonus to constructing color-dual numerators, the BCJ
double copy allowed us to construct closed-string-insertion numerators
from pairings of the open-string-insertion numerators.  We found that
regardless of the prescription used, the double-copy numerators
produced the correct kinematic structures to match against directly
calculated cuts, but not the correct normalziation.  In fact, the
mismatch in normalization is explained by observaing that the sum over
all allowed prescriptions (with degeneracy) exactly matches the
directly calculated cuts.  This sum including degeneracy is in direct
analogy with terms and normalizations in the KLT representation of the
tree-level closed-string amplitudes, and may suggest a similar
decomposition for the one-loop numerators.

\section*{Acknowledgements}
We give significant thanks Henrik Johansson and Oliver Schlotterer for
discussion and feedback throughout the project.  We appreciate Oliver
Schlotterer and Fei Teng for comments on the manuscript.  This work is
an extension of MT's master's project at Uppsala University.  AE is
supported in part by the Knut and Alice Wallenberg Foundation under
KAW 2018.0116, {\it From Scattering Amplitudes to Gravitational
  Waves}, by the USDOE under contract DE-SC0015910, and by
Northwestern University via the Amplitudes and Insight Group,
Department of Physics and Astronomy, and Weinberg College of Arts and
Sciences.  The diagrams in this manuscript were made using
TikZ-Feynman \cite{Ellis:2016jkw}.

\bibliographystyle{JHEP}
\bibliography{zetas_bcj_bib}

\providecommand{\href}[2]{#2}\begingroup\raggedright\begin{thebibliography}{10}

\bibitem{Kawai:1985xq}
H.~Kawai, D.~C. Lewellen and S.~H.~H. Tye, \emph{{A Relation Between Tree
  Amplitudes of Closed and Open Strings}},
  \href{http://dx.doi.org/10.1016/0550-3213(86)90362-7}{\emph{Nucl. Phys. B}
  {\bf 269} (1986) 1--23}.

\bibitem{Bern:2010ue}
Z.~Bern, J.~J.~M. Carrasco and H.~Johansson, \emph{{Perturbative Quantum
  Gravity as a Double Copy of Gauge Theory}},
  \href{http://dx.doi.org/10.1103/PhysRevLett.105.061602}{\emph{Phys.Rev.Lett.}
  {\bf 105} (2010) }, [\href{http://arxiv.org/abs/1004.0476}{{\tt 1004.0476}}].

\bibitem{Cachazo:2013iea}
F.~Cachazo, S.~He and E.~Y. Yuan, \emph{{Scattering of Massless Particles:
  Scalars, Gluons and Gravitons}},
  \href{http://dx.doi.org/10.1007/JHEP07(2014)033}{\emph{JHEP} {\bf 07} (2014)
  033}, [\href{http://arxiv.org/abs/1309.0885}{{\tt 1309.0885}}].

\bibitem{Bern:2019prr}
Z.~Bern, J.~J. Carrasco, M.~Chiodaroli, H.~Johansson and R.~Roiban, \emph{{The
  Duality Between Color and Kinematics and its Applications}},
  \href{http://arxiv.org/abs/1909.01358}{{\tt 1909.01358}}.

\bibitem{Chi:2021mio}
H.-H. Chi, H.~Elvang, A.~Herderschee, C.~R.~T. Jones and S.~Paranjape,
  \emph{{Generalizations of the double-copy: the KLT bootstrap}},
  \href{http://dx.doi.org/10.1007/JHEP03(2022)077}{\emph{JHEP} {\bf 03} (2022)
  077}, [\href{http://arxiv.org/abs/2106.12600}{{\tt 2106.12600}}].

\bibitem{Carrasco:2016ldy}
J.~J.~M. Carrasco, C.~R. Mafra and O.~Schlotterer, \emph{{Abelian Z-theory:
  NLSM amplitudes and $\alpha$'-corrections from the open string}},
  \href{http://dx.doi.org/10.1007/JHEP06(2017)093}{\emph{JHEP} {\bf 06} (2017)
  093}, [\href{http://arxiv.org/abs/1608.02569}{{\tt 1608.02569}}].

\bibitem{Carrasco:2016ygv}
J.~J.~M. Carrasco, C.~R. Mafra and O.~Schlotterer, \emph{{Semi-abelian
  Z-theory: NLSM$+\phi^{3}$ from the open string}},
  \href{http://dx.doi.org/10.1007/JHEP08(2017)135}{\emph{JHEP} {\bf 08} (2017)
  135}, [\href{http://arxiv.org/abs/1612.06446}{{\tt 1612.06446}}].

\bibitem{Schlotterer:2012ny}
O.~Schlotterer and S.~Stieberger, \emph{{Motivic Multiple Zeta Values and
  Superstring Amplitudes}},
  \href{http://dx.doi.org/10.1088/1751-8113/46/47/475401}{\emph{J. Phys. A}
  {\bf 46} (2013) 475401}, [\href{http://arxiv.org/abs/1205.1516}{{\tt
  1205.1516}}].

\bibitem{Stieberger:2013wea}
S.~Stieberger, \emph{{Closed superstring amplitudes, single-valued multiple
  zeta values and the Deligne associator}},
  \href{http://dx.doi.org/10.1088/1751-8113/47/15/155401}{\emph{J.Phys.A} {\bf
  47} (2014) }, [\href{http://arxiv.org/abs/1310.3259}{{\tt 1310.3259}}].

\bibitem{Stieberger:2014hba}
S.~Stieberger and T.~R. Taylor, \emph{{Closed String Amplitudes as
  Single-Valued Open String Amplitudes}},
  \href{http://dx.doi.org/10.1016/j.nuclphysb.2014.02.005}{\emph{Nucl.Phys.B}
  {\bf 881} (2014) 269}, [\href{http://arxiv.org/abs/1401.1218}{{\tt
  1401.1218}}].

\bibitem{Schlotterer:2018zce}
O.~Schlotterer and O.~Schnetz, \emph{{Closed strings as single-valued open
  strings: A genus-zero derivation}},
  \href{http://dx.doi.org/10.1088/1751-8121/aaea14}{\emph{J. Phys. A} {\bf 52}
  (2019) 045401}, [\href{http://arxiv.org/abs/1808.00713}{{\tt 1808.00713}}].

\bibitem{Vanhove:2018elu}
P.~Vanhove and F.~Zerbini, \emph{{Single-valued hyperlogarithms, correlation
  functions and closed string amplitudes}},
  \href{http://arxiv.org/abs/1812.03018}{{\tt 1812.03018}}.

\bibitem{Brown:2019wna}
F.~Brown and C.~Dupont, \emph{{Single-valued integration and superstring
  amplitudes in genus zero}},
  \href{http://dx.doi.org/10.1007/s00220-021-03969-4}{\emph{Commun.Math.Phys.}
  {\bf 382} (2021) 815}, [\href{http://arxiv.org/abs/1910.01107}{{\tt
  1910.01107}}].

\bibitem{Edison:2021ebi}
A.~Edison, M.~Guillen, H.~Johansson, O.~Schlotterer and F.~Teng,
  \emph{{One-loop matrix elements of effective superstring interactions:
  \ensuremath{\alpha}'-expanding loop integrands}},
  \href{http://dx.doi.org/10.1007/JHEP12(2021)007}{\emph{JHEP} {\bf 12} (2021)
  007}, [\href{http://arxiv.org/abs/2107.08009}{{\tt 2107.08009}}].

\bibitem{DHoker:2019blr}
E.~D'Hoker and M.~B. Green, \emph{{Exploring transcendentality in superstring
  amplitudes}}, \href{http://dx.doi.org/10.1007/JHEP07(2019)149}{\emph{JHEP}
  {\bf 07} (2019) 149}, [\href{http://arxiv.org/abs/1906.01652}{{\tt
  1906.01652}}].

\bibitem{Bern:2007ct}
Z.~Bern, J.~Carrasco, H.~Johansson and D.~Kosower, \emph{{Maximally
  supersymmetric planar Yang-Mills amplitudes at five loops}},
  \href{http://dx.doi.org/10.1103/PhysRevD.76.125020}{\emph{Phys.Rev.D} {\bf
  76} (2007) }, [\href{http://arxiv.org/abs/0705.1864}{{\tt 0705.1864}}].

\bibitem{Bern:2008pv}
Z.~Bern, J.~Carrasco, L.~J. Dixon, H.~Johansson and R.~Roiban, \emph{{Manifest
  Ultraviolet Behavior for the Three-Loop Four-Point Amplitude of N=8
  Supergravity}},
  \href{http://dx.doi.org/10.1103/PhysRevD.78.105019}{\emph{Phys.Rev.D} {\bf
  78} (2008) }, [\href{http://arxiv.org/abs/0808.4112}{{\tt 0808.4112}}].

\bibitem{Bern:2010tq}
Z.~Bern, J.~Carrasco, L.~J. Dixon, H.~Johansson and R.~Roiban, \emph{{The
  Complete Four-Loop Four-Point Amplitude in N=4 Super-Yang-Mills Theory}},
  \href{http://dx.doi.org/10.1103/PhysRevD.82.125040}{\emph{Phys.Rev.D} {\bf
  82} (2010) }, [\href{http://arxiv.org/abs/1008.3327}{{\tt 1008.3327}}].

\bibitem{Bern:2012uf}
Z.~Bern, J.~Carrasco, L.~Dixon, H.~Johansson and R.~Roiban, \emph{{Simplifying
  Multiloop Integrands and Ultraviolet Divergences of Gauge Theory and Gravity
  Amplitudes}},
  \href{http://dx.doi.org/10.1103/PhysRevD.85.105014}{\emph{Phys.Rev.D} {\bf
  85} (2012) }, [\href{http://arxiv.org/abs/1201.5366}{{\tt 1201.5366}}].

\bibitem{Bern:2018jmv}
Z.~Bern, J.~J. Carrasco, W.-M. Chen, A.~Edison, H.~Johansson, J.~Parra-Martinez
  et~al., \emph{{Ultraviolet Properties of $\mathcal N = 8$ Supergravity at
  Five Loops}},
  \href{http://dx.doi.org/10.1103/PhysRevD.98.086021}{\emph{Phys.Rev.D} {\bf
  98} (2018) }, [\href{http://arxiv.org/abs/1804.09311}{{\tt 1804.09311}}].

\bibitem{Mangano:1990by}
M.~L. Mangano and S.~J. Parke, \emph{{Multiparton amplitudes in gauge
  theories}},
  \href{http://dx.doi.org/10.1016/0370-1573(91)90091-Y}{\emph{Phys.Rept.} {\bf
  200} (1991) 301}, [\href{http://arxiv.org/abs/hep-th/0509223}{{\tt
  hep-th/0509223}}].

\bibitem{Zeppenfeld:1988bz}
D.~Zeppenfeld, \emph{{Diagonalization of Color Factors}},
  \href{http://dx.doi.org/10.1142/S0217751X88000916}{\emph{Int.J.Mod.Phys.A}
  {\bf 3} (1988) 2175}.

\bibitem{Mangano:1988kk}
M.~L. Mangano, \emph{{The Color Structure of Gluon Emission}},
  \href{http://dx.doi.org/10.1016/0550-3213(88)90453-1}{\emph{Nucl.Phys.B} {\bf
  309} (1988) 461}.

\bibitem{Kosower:1987ic}
D.~Kosower, B.-H. Lee and V.~Nair, \emph{{MULTI GLUON SCATTERING: A STRING
  BASED CALCULATION}},
  \href{http://dx.doi.org/10.1016/0370-2693(88)90085-8}{\emph{Phys.Lett.B} {\bf
  201} (1988) 85}.

\bibitem{Berends:1987cv}
F.~A. Berends and W.~Giele, \emph{{The Six Gluon Process as an Example of
  Weyl-Van Der Waerden Spinor Calculus}},
  \href{http://dx.doi.org/10.1016/0550-3213(87)90604-3}{\emph{Nucl.Phys.B} {\bf
  294} (1987) 700}.

\bibitem{Cvitanovic:1980bu}
P.~Cvitanovic, P.~Lauwers and P.~Scharbach, \emph{{Gauge Invariance Structure
  of Quantum Chromodynamics}},
  \href{http://dx.doi.org/10.1016/0550-3213(81)90098-5}{\emph{Nucl.Phys.B} {\bf
  186} (1981) 165}.

\bibitem{Kleiss:1988ne}
R.~Kleiss and H.~Kuijf, \emph{{Multi - Gluon Cross-sections and Five Jet
  Production at Hadron Colliders}},
  \href{http://dx.doi.org/10.1016/0550-3213(89)90574-9}{\emph{Nucl.Phys.B} {\bf
  312} (1989) 616}.

\bibitem{Bern:2008qj}
Z.~Bern, J.~Carrasco and H.~Johansson, \emph{{New Relations for Gauge-Theory
  Amplitudes}},
  \href{http://dx.doi.org/10.1103/PhysRevD.78.085011}{\emph{Phys.Rev.D} {\bf
  78} (2008) }, [\href{http://arxiv.org/abs/0805.3993}{{\tt 0805.3993}}].

\bibitem{Bern:2010yg}
Z.~Bern, T.~Dennen, Y.-t. Huang and M.~Kiermaier, \emph{{Gravity as the Square
  of Gauge Theory}},
  \href{http://dx.doi.org/10.1103/PhysRevD.82.065003}{\emph{Phys.Rev.D} {\bf
  82} (2010) }, [\href{http://arxiv.org/abs/1004.0693}{{\tt 1004.0693}}].

\bibitem{Cachazo:2013hca}
F.~Cachazo, S.~He and E.~Y. Yuan, \emph{{Scattering of Massless Particles in
  Arbitrary Dimensions}},
  \href{http://dx.doi.org/10.1103/PhysRevLett.113.171601}{\emph{Phys.Rev.Lett.}
  {\bf 113} (2014) }, [\href{http://arxiv.org/abs/1307.2199}{{\tt 1307.2199}}].

\bibitem{Lam:2016tlk}
C.~Lam and Y.-P. Yao, \emph{{Evaluation of the Cachazo-He-Yuan gauge
  amplitude}},
  \href{http://dx.doi.org/10.1103/PhysRevD.93.105008}{\emph{Phys.Rev.D} {\bf
  93} (2016) }, [\href{http://arxiv.org/abs/1602.06419}{{\tt 1602.06419}}].

\bibitem{Du:2017gnh}
Y.-J. Du, B.~Feng and F.~Teng, \emph{{Expansion of All Multitrace Tree Level
  EYM Amplitudes}},
  \href{http://dx.doi.org/10.1007/JHEP12(2017)038}{\emph{JHEP} {\bf 12} (2017)
  038}, [\href{http://arxiv.org/abs/1708.04514}{{\tt 1708.04514}}].

\bibitem{Fu:2017uzt}
C.-H. Fu, Y.-J. Du, R.~Huang and B.~Feng, \emph{{Expansion of
  Einstein-Yang-Mills Amplitude}},
  \href{http://dx.doi.org/10.1007/JHEP09(2017)021}{\emph{JHEP} {\bf 09} (2017)
  021}, [\href{http://arxiv.org/abs/1702.08158}{{\tt 1702.08158}}].

\bibitem{Du:2017kpo}
Y.-J. Du and F.~Teng, \emph{{BCJ numerators from reduced Pfaffian}},
  \href{http://dx.doi.org/10.1007/JHEP04(2017)033}{\emph{JHEP} {\bf 04} (2017)
  033}, [\href{http://arxiv.org/abs/1703.05717}{{\tt 1703.05717}}].

\bibitem{Edison:2020ehu}
A.~Edison and F.~Teng, \emph{{Efficient Calculation of Crossing Symmetric BCJ
  Tree Numerators}},
  \href{http://dx.doi.org/10.1007/JHEP12(2020)138}{\emph{JHEP} {\bf 12} (2020)
  138}, [\href{http://arxiv.org/abs/2005.03638}{{\tt 2005.03638}}].

\bibitem{Cheung:2021zvb}
C.~Cheung and J.~Mangan, \emph{{Covariant color-kinematics duality}},
  \href{http://dx.doi.org/10.1007/JHEP11(2021)069}{\emph{JHEP} {\bf 11} (2021)
  069}, [\href{http://arxiv.org/abs/2108.02276}{{\tt 2108.02276}}].

\bibitem{Ben-Shahar:2021zww}
M.~Ben-Shahar and H.~Johansson, \emph{{Off-shell color-kinematics duality for
  Chern-Simons}}, \href{http://dx.doi.org/10.1007/JHEP08(2022)035}{\emph{JHEP}
  {\bf 08} (2022) 035}, [\href{http://arxiv.org/abs/2112.11452}{{\tt
  2112.11452}}].

\bibitem{Carrasco:2021ptp}
J.~J.~M. Carrasco, L.~Rodina and S.~Zekioglu, \emph{{Composing effective
  prediction at five points}},
  \href{http://dx.doi.org/10.1007/JHEP06(2021)169}{\emph{JHEP} {\bf 06} (2021)
  169}, [\href{http://arxiv.org/abs/2104.08370}{{\tt 2104.08370}}].

\bibitem{Carrasco:2019yyn}
J.~J.~M. Carrasco, L.~Rodina, Z.~Yin and S.~Zekioglu, \emph{{Simple encoding of
  higher derivative gauge and gravity counterterms}},
  \href{http://dx.doi.org/10.1103/PhysRevLett.125.251602}{\emph{Phys. Rev.
  Lett.} {\bf 125} (2020) 251602}, [\href{http://arxiv.org/abs/1910.12850}{{\tt
  1910.12850}}].

\bibitem{Minahan:1987ha}
J.~A. Minahan, \emph{{One Loop Amplitudes on Orbifolds and the Renormalization
  of Coupling Constants}},
  \href{http://dx.doi.org/10.1016/0550-3213(88)90303-3}{\emph{Nucl.Phys.B} {\bf
  298} (1988) 36}.

\bibitem{Edison:2022jln}
A.~Edison, S.~He, H.~Johansson, O.~Schlotterer, F.~Teng and Y.~Zhang,
  \emph{{Perfecting one-loop BCJ numerators in SYM and supergravity}},
  \href{http://dx.doi.org/10.1007/JHEP02(2023)164}{\emph{JHEP} {\bf 02} (2023)
  164}, [\href{http://arxiv.org/abs/2211.00638}{{\tt 2211.00638}}].

\bibitem{Bern:2017ucb}
Z.~Bern, J.~J.~M. Carrasco, W.-M. Chen, H.~Johansson, R.~Roiban and M.~Zeng,
  \emph{{Five-loop four-point integrand of $N=8$ supergravity as a generalized
  double copy}},
  \href{http://dx.doi.org/10.1103/PhysRevD.96.126012}{\emph{Phys. Rev. D} {\bf
  96} (2017) 126012}, [\href{http://arxiv.org/abs/1708.06807}{{\tt
  1708.06807}}].

\bibitem{Bern:2013yya}
Z.~Bern, S.~Davies, T.~Dennen, Y.-t. Huang and J.~Nohle,
  \emph{{Color-Kinematics Duality for Pure Yang-Mills and Gravity at One and
  Two Loops}}, \href{http://dx.doi.org/10.1103/PhysRevD.92.045041}{\emph{Phys.
  Rev. D} {\bf 92} (2015) 045041}, [\href{http://arxiv.org/abs/1303.6605}{{\tt
  1303.6605}}].

\bibitem{Bern:2015ooa}
Z.~Bern, S.~Davies and J.~Nohle, \emph{{Double-Copy Constructions and Unitarity
  Cuts}}, \href{http://dx.doi.org/10.1103/PhysRevD.93.105015}{\emph{Phys. Rev.
  D} {\bf 93} (2016) 105015}, [\href{http://arxiv.org/abs/1510.03448}{{\tt
  1510.03448}}].

\bibitem{Mogull:2015adi}
G.~Mogull and D.~O'Connell, \emph{{Overcoming Obstacles to Colour-Kinematics
  Duality at Two Loops}},
  \href{http://dx.doi.org/10.1007/JHEP12(2015)135}{\emph{JHEP} {\bf 12} (2015)
  135}, [\href{http://arxiv.org/abs/1511.06652}{{\tt 1511.06652}}].

\bibitem{Johansson:2017bfl}
H.~Johansson, G.~K\"alin and G.~Mogull, \emph{{Two-loop supersymmetric QCD and
  half-maximal supergravity amplitudes}},
  \href{http://dx.doi.org/10.1007/JHEP09(2017)019}{\emph{JHEP} {\bf 09} (2017)
  019}, [\href{http://arxiv.org/abs/1706.09381}{{\tt 1706.09381}}].

\bibitem{Bridges:2021ebs}
E.~Bridges and C.~R. Mafra, \emph{{Local BCJ numerators for ten-dimensional SYM
  at one loop}}, \href{http://dx.doi.org/10.1007/JHEP07(2021)031}{\emph{JHEP}
  {\bf 07} (2021) 031}, [\href{http://arxiv.org/abs/2102.12943}{{\tt
  2102.12943}}].

\bibitem{Vaman:2010ez}
D.~Vaman and Y.-P. Yao, \emph{{Constraints and Generalized Gauge
  Transformations on Tree-Level Gluon and Graviton Amplitudes}},
  \href{http://dx.doi.org/10.1007/JHEP11(2010)028}{\emph{JHEP} {\bf 11} (2010)
  028}, [\href{http://arxiv.org/abs/1007.3475}{{\tt 1007.3475}}].

\bibitem{Du:2011js}
Y.-J. Du, B.~Feng and C.-H. Fu, \emph{{BCJ Relation of Color Scalar Theory and
  KLT Relation of Gauge Theory}},
  \href{http://dx.doi.org/10.1007/JHEP08(2011)129}{\emph{JHEP} {\bf 08} (2011)
  129}, [\href{http://arxiv.org/abs/1105.3503}{{\tt 1105.3503}}].

\bibitem{Bjerrum-Bohr:2012kaa}
N.~Bjerrum-Bohr, P.~H. Damgaard, R.~Monteiro and D.~O'Connell, \emph{{Algebras
  for Amplitudes}},
  \href{http://dx.doi.org/10.1007/JHEP06(2012)061}{\emph{JHEP} {\bf 06} (2012)
  061}, [\href{http://arxiv.org/abs/1203.0944}{{\tt 1203.0944}}].

\bibitem{DelDuca:1999rs}
V.~Del~Duca, L.~J. Dixon and F.~Maltoni, \emph{{New color decompositions for
  gauge amplitudes at tree and loop level}},
  \href{http://dx.doi.org/10.1016/S0550-3213(99)00809-3}{\emph{Nucl. Phys. B}
  {\bf 571} (2000) 51--70}, [\href{http://arxiv.org/abs/hep-ph/9910563}{{\tt
  hep-ph/9910563}}].

\bibitem{Mafra:2015vca}
C.~R. Mafra and O.~Schlotterer, \emph{{Berends-Giele recursions and the BCJ
  duality in superspace and components}},
  \href{http://dx.doi.org/10.1007/JHEP03(2016)097}{\emph{JHEP} {\bf 03} (2016)
  097}, [\href{http://arxiv.org/abs/1510.08846}{{\tt 1510.08846}}].

\bibitem{Bjerrum-Bohr:2016axv}
N.~E.~J. Bjerrum-Bohr, J.~L. Bourjaily, P.~H. Damgaard and B.~Feng,
  \emph{{Manifesting Color-Kinematics Duality in the Scattering Equation
  Formalism}}, \href{http://dx.doi.org/10.1007/JHEP09(2016)094}{\emph{JHEP}
  {\bf 09} (2016) 094}, [\href{http://arxiv.org/abs/1608.00006}{{\tt
  1608.00006}}].

\bibitem{Edison:2020uzf}
A.~Edison, S.~He, O.~Schlotterer and F.~Teng, \emph{{One-loop Correlators and
  BCJ Numerators from Forward Limits}},
  \href{http://dx.doi.org/10.1007/JHEP09(2020)079}{\emph{JHEP} {\bf 09} (2020)
  079}, [\href{http://arxiv.org/abs/2005.03639}{{\tt 2005.03639}}].

\bibitem{Bern:1998sv}
Z.~Bern, L.~J. Dixon, M.~Perelstein and J.~Rozowsky, \emph{{Multileg one loop
  gravity amplitudes from gauge theory}},
  \href{http://dx.doi.org/10.1016/S0550-3213(99)00029-2}{\emph{Nucl.Phys.B}
  {\bf 546} (1999) 423}, [\href{http://arxiv.org/abs/hep-th/9811140}{{\tt
  hep-th/9811140}}].

\bibitem{Bjerrum-Bohr:2010pnr}
N.~Bjerrum-Bohr, P.~H. Damgaard, T.~Sondergaard and P.~Vanhove, \emph{{The
  Momentum Kernel of Gauge and Gravity Theories}},
  \href{http://dx.doi.org/10.1007/JHEP01(2011)001}{\emph{JHEP} {\bf 01} (2011)
  001}, [\href{http://arxiv.org/abs/1010.3933}{{\tt 1010.3933}}].

\bibitem{Mafra:2011nv}
C.~R. Mafra, O.~Schlotterer and S.~Stieberger, \emph{{Complete N-Point
  Superstring Disk Amplitude I. Pure Spinor Computation}},
  \href{http://dx.doi.org/10.1016/j.nuclphysb.2013.04.023}{\emph{Nucl.Phys.B}
  {\bf 873} (2013) 419}, [\href{http://arxiv.org/abs/1106.2645}{{\tt
  1106.2645}}].

\bibitem{Broedel:2013tta}
J.~Broedel, O.~Schlotterer and S.~Stieberger, \emph{{Polylogarithms, Multiple
  Zeta Values and Superstring Amplitudes}},
  \href{http://dx.doi.org/10.1002/prop.201300019}{\emph{Fortsch.Phys.} {\bf 61}
  (2013) 812}, [\href{http://arxiv.org/abs/1304.7267}{{\tt 1304.7267}}].

\bibitem{BGap}
C.~Mafra and O.~Schlotterer. \url{https://repo.or.cz/BGap.git}.

\bibitem{MZVWebsite}
J.~Broedel, O.~Schlotterer and S.~Stieberger.
  \url{https://wwwth.mpp.mpg.de/members/stieberg/mzv/}.

\bibitem{Broedel:2013aza}
J.~Broedel, O.~Schlotterer, S.~Stieberger and T.~Terasoma, \emph{{All order
  $\alpha^{\prime}$-expansion of superstring trees from the Drinfeld
  associator}}, \href{http://dx.doi.org/10.1103/PhysRevD.89.066014}{\emph{Phys.
  Rev. D} {\bf 89} (2014) 066014}, [\href{http://arxiv.org/abs/1304.7304}{{\tt
  1304.7304}}].

\bibitem{Mafra:2016mcc}
C.~R. Mafra and O.~Schlotterer, \emph{{Non-abelian $Z$-theory: Berends-Giele
  recursion for the $\alpha'$-expansion of disk integrals}},
  \href{http://dx.doi.org/10.1007/JHEP01(2017)031}{\emph{JHEP} {\bf 01} (2017)
  031}, [\href{http://arxiv.org/abs/1609.07078}{{\tt 1609.07078}}].

\bibitem{Kitazawa:1987xj}
Y.~Kitazawa, \emph{{Effective Lagrangian for Open Superstring From Five Point
  Function}},
  \href{http://dx.doi.org/10.1016/0550-3213(87)90396-8}{\emph{Nucl.Phys.B} {\bf
  289} (1987) 599}.

\bibitem{Barreiro:2005hv}
L.~A. Barreiro and R.~Medina, \emph{{5-field terms in the open superstring
  effective action}},
  \href{http://dx.doi.org/10.1088/1126-6708/2005/03/055}{\emph{JHEP} {\bf 03}
  (2005) 055}, [\href{http://arxiv.org/abs/hep-th/0503182}{{\tt
  hep-th/0503182}}].

\bibitem{Barreiro:2012aw}
L.~A. Barreiro and R.~Medina, \emph{{Revisiting the S-matrix approach to the
  open superstring low energy effective lagrangian}},
  \href{http://dx.doi.org/10.1007/JHEP10(2012)108}{\emph{JHEP} {\bf 10} (2012)
  108}, [\href{http://arxiv.org/abs/1208.6066}{{\tt 1208.6066}}].

\bibitem{Koerber:2002zb}
P.~Koerber and A.~Sevrin, \emph{{The NonAbelian D-brane effective action
  through order alpha-prime**4}},
  \href{http://dx.doi.org/10.1088/1126-6708/2002/10/046}{\emph{JHEP} {\bf 10}
  (2002) 046}, [\href{http://arxiv.org/abs/hep-th/0208044}{{\tt
  hep-th/0208044}}].

\bibitem{Oprisa:2005wu}
D.~Oprisa and S.~Stieberger, \emph{{Six gluon open superstring disk amplitude,
  multiple hypergeometric series and Euler-Zagier sums}},
  \href{http://arxiv.org/abs/hep-th/0509042}{{\tt hep-th/0509042}}.

\bibitem{Schnetz:2013hqa}
O.~Schnetz, \emph{{Graphical functions and single-valued multiple
  polylogarithms}},
  \href{http://dx.doi.org/10.4310/CNTP.2014.v8.n4.a1}{\emph{Commun.Num.Theor.Phys.}
  {\bf 08} (2014) 589}, [\href{http://arxiv.org/abs/1302.6445}{{\tt
  1302.6445}}].

\bibitem{Brown:2013gia}
F.~Brown, \emph{{Single-valued Motivic Periods and Multiple Zeta Values}},
  \href{http://dx.doi.org/10.1017/fms.2014.18}{\emph{SIGMA} {\bf 2} (2014)
  e25}, [\href{http://arxiv.org/abs/1309.5309}{{\tt 1309.5309}}].

\bibitem{Roehrig:2017gbt}
K.~A. Roehrig and D.~Skinner, \emph{{A Gluing Operator for the Ambitwistor
  String}}, \href{http://dx.doi.org/10.1007/JHEP01(2018)069}{\emph{JHEP} {\bf
  01} (2018) 069}, [\href{http://arxiv.org/abs/1709.03262}{{\tt 1709.03262}}].

\bibitem{Kosmopoulos:2020pcd}
D.~Kosmopoulos, \emph{{Simplifying D-dimensional physical-state sums in gauge
  theory and gravity}},
  \href{http://dx.doi.org/10.1103/PhysRevD.105.056025}{\emph{Phys.Rev.D} {\bf
  105} (2022) }, [\href{http://arxiv.org/abs/2009.00141}{{\tt 2009.00141}}].

\bibitem{Elvang:2015rqa}
H.~Elvang and Y.-t. Huang, \emph{{Scattering Amplitudes in Gauge Theory and
  Gravity}}.
\newblock Cambridge University Press, 4, 2015.

\bibitem{Bjerrum-Bohr:2009ulz}
N.~E.~J. Bjerrum-Bohr, P.~H. Damgaard and P.~Vanhove, \emph{{Minimal Basis for
  Gauge Theory Amplitudes}},
  \href{http://dx.doi.org/10.1103/PhysRevLett.103.161602}{\emph{Phys. Rev.
  Lett.} {\bf 103} (2009) 161602}, [\href{http://arxiv.org/abs/0907.1425}{{\tt
  0907.1425}}].

\bibitem{Stieberger:2009hq}
S.~Stieberger, \emph{{Open \& Closed vs. Pure Open String Disk Amplitudes}},
  \href{http://arxiv.org/abs/0907.2211}{{\tt 0907.2211}}.

\bibitem{Broedel:2012rc}
J.~Broedel and L.~J. Dixon, \emph{{Color-kinematics duality and double-copy
  construction for amplitudes from higher-dimension operators}},
  \href{http://dx.doi.org/10.1007/JHEP10(2012)091}{\emph{JHEP} {\bf 10} (2012)
  091}, [\href{http://arxiv.org/abs/1208.0876}{{\tt 1208.0876}}].

\bibitem{Ellis:2016jkw}
J.~Ellis, \emph{{TikZ-Feynman: Feynman diagrams with TikZ}},
  \href{http://dx.doi.org/10.1016/j.cpc.2016.08.019}{\emph{Comput. Phys.
  Commun.} {\bf 210} (2017) 103--123},
  [\href{http://arxiv.org/abs/1601.05437}{{\tt 1601.05437}}].

\end{thebibliography}\endgroup

\end{document}